# Lattice-based simulation of the effects of nutrient concentration and magnetic field exposure on yeast colony growth and morphology

Rebekah Hall[1] and Daniel A. Charlebois[2,3,*]
[1]Department of Mathematical and Statistical Sciences, University of Alberta, Edmonton, AB, T6G-2G1, Canada
[2]Department of Physics, University of Alberta, Edmonton, AB, T6G-2E1, Canada
[3]Department of Biological Sciences, University of Alberta, Edmonton, AB, T6G-2E9, Canada
*Corresponding Author: dcharleb@ualberta.ca

## Abstract

Yeasts exist in communities that expand over space and time to form complex structures and patterns. We developed a computational lattice-based framework to perform spatial-temporal simulations of budding yeast colonies exposed to different nutrient and magnetic field conditions. The budding patterns of haploid and diploid yeast cells were incorporated into the framework, as well as the filamentous growth that occurs in yeast colonies under nutrient limiting conditions. Simulation of the lattice-based model predicted that magnetic fields decrease colony growth rate, density, and roundness. Magnetic field simulations further predicted that colony elongation and boundary fluctuations increase in a nutrient- and ploidy-dependent manner. These *in-silico* predictions are an important step towards understanding the effects of the physico-chemical environment on microbial colonies and for informing bioelectromagnetic experiments on yeast colony biofilms and fungal pathogens.

## INTRODUCTION

Proliferating cells that interact with their environment can lead to the formation of complex multicellular structures and biological patterns. For instance, the budding yeast *Saccharomyces cerevisiae* forms colonies, which are organized microbial communities [1] that can display intricate multicellular patterns [2]. Budding yeast can also exit as spatially and metabolically structured communities embedded in an extracellular polymer matrix known as a biofilm [3, 4]. Though biofilm formation in eukaryotes is not well understood [2], biofilms are known to adhere to medical devices [4, 5] and to increase the resistance to antimicrobial drugs [6]. Yeast communities respond to their environment, such as the concentration of nutrients and the physical properties of the growth substrate [2], as well as to cell-to-cell interactions [3]. Improving our fundamental understanding and ability to quantitatively predict the effects of the physico-chemical environment on the growth and structure of microbial communities will be important for designing new biomaterials [7] and for mitigating antimicrobial resistance [4, 6].

The budding yeast *Saccharomyces cerevisiae* has been used as a model organism to study the effects of electromagnetic fields (EMFs). A range of effects from EMFs have been observed in *S. cerevisiae*, including altered gene expression [8, 9], decreased cell viability and growth rates [10, 11], and frequency-dependent proliferative responses [12]. Egami et al. experimentally investigated the changes in the budding angle and the size of individual *S. cerevisiae* cells exposed to static magnetic fields (MFs) [13]. Though the effect of MF exposure on cell growth was minor, the budding direction of the daughter yeast cells was significantly affected by homogeneous and inhomogeneous MFs. The budding angle of the daughter cells was found to be mainly oriented in the direction of the homogeneous MF, whereas cells tended to bud perpendicular to the direction of the inhomogeneous MF in regions where the MF gradient was high. These experiments provide single-cell data that can be used to model the emergent properties of yeast colonies and colony biofilms under MF exposure.

Agent-based models, including lattice-based models, have been used to simulate mechanical and physiological phenomena in cells and tissues [14]. In contrast to a continuum model, an agent-based model treats cells as separate units, which provides an ideal framework for investigating the effects of spatial inhomogeneities and phenotypic variability between cells, as well as the collective impact of cellular responses, on population dynamics. Lattice-based approaches are ubiquitous in the field of physics. One well-known application is the Ising model of ferromagnetism in statistical mechanics [15]. In biophysics, protein folding is commonly described by a class of lattice models of compact polymers in which the constituent amino acids are constrained to occupy a regular array of positions in space [16, 17]. Lattice-based models have also been used to model cellular proliferation and migration, as well as pattern formation [18, 19]. While lattice-based approaches have been used to spatially and temporally model cell populations [20, 21], they have not been used to predict the effects of EMFs on cell populations.

Here we present a computational 2D lattice-based framework to investigate the effects of nutrient and magnetic field conditions on expanding colonies of yeast cells. The ploidy and corresponding budding patterns [22], as well as magnetic field dependent budding angles [13], of yeast cells are incorporated into the model. Simulations of the lattice-based model are in agreement with known experimental nutrient-dependent growth and morphological effects [2], as well as magnetic-field dependent effects on single-cell budding angle distributions [13]. The modeling framework predicts how the exposure of budding yeast cells to magnetic fields affects colony growth and morphology. Magnetic fields are found to decrease colony growth, density, and roundness, as well as to increase colony elongation and boundary fluctuations in a nutrient- and ploidy-dependent manner.

## SIMULATION FRAMEWORK

### *Standard Culture Conditions*

We began by developing a 2D lattice-based model to simulate yeast colony expansion under standard laboratory conditions (Fig. 1). That is, yeast cells growing on nutrient rich (such as YPD medium) agar plates in an incubator at 30°C with no applied magnetic fields [23]. We made the standard assumption that all yeast cells in a colony are generated from a single cell and are genetically identical.

When a budding yeast cell prepares for division, it becomes polarized along the mother-bud axis, which determines the direction that the yeast cell will bud [22, 24]. Bud-site selection generally follows one of two patterns: either an axial budding pattern (the primary pattern for haploid yeast cells; Fig. 2A) or the bipolar budding pattern (the primary budding pattern for diploid yeast cells; Fig. 2B). Regardless of the budding pattern, the budding direction is determined with respect to the direction of budding in the previous replication, which is marked by a bud scar. If a cell has never replicated before, the scar will be at the site at which it separated from its mother cell and is referred to as a birth scar. An axial budding pattern requires that the yeast bud grows adjacent to the bud scar. A bipolar budding pattern is more complex. The daughter cell's first bud will form at the end furthest from its birth scar. The mother cell's next bud will form at the end opposite its previous daughter cell, or it may bud at the end closest to its bud scar. Though these are the predominant bipolar patterns occasionally the daughter cell may also bud adjacent to its bud scar [22, 25].

In our computational model, the bud (or birth) scar of each cell is saved as a property of this object and represented by a vector, [x, y]. This vector indicates the direction of polarization by the difference in coordinates between the cell and its previous bud (or in the case of a daughter's birth scar, its mother cell). If the coordinates of a cell are (x, y) and it produces a bud at coordinates (x', y'), then the cell's bud scar is given by [x-x', y-y'] and the new bud's birth scar will be the opposite, [x'-x, y'-y]. This model uses a "Moore neighborhood", meaning that all eight lattice sites surrounding a cell are considered possible sites into which a cell can bud [26]. If all eight sites are occupied by cells, the surrounding area is considered too crowded and the cell cannot bud.

To model axial budding, the angles between a cell's bud scar and each empty lattice site surrounding it are calculated (Fig. 2A). We observed that if the smallest angle is chosen as the direction of budding, and in the case where one of two empty sites with the same smallest angle is chosen randomly, that diamond-shaped colonies are produced (data not shown); diamond-shaped colonies do not resemble the circular colonies commonly grown on agar plates in the laboratory [23]. To rectify this, budding is restricted to the two sites 45° from the bud scar (Fig. 2A). If both adjacent sites are empty, one site is chosen randomly. If only one site is empty, the cell will bud into the empty site. If both adjacent sites are occupied, the cell will randomly bud into one of the two sites, pushing the occupying cell into a randomly selected empty site nearby. If there are no empty sites around the occupying cell, then the area is considered too crowded and the cell will not be able to bud. The resulting simulations yielded approximately circular colonies in rich-nutrient conditions and in the absence of MFs (Fig. 3), in agreement with standard experimental observations (e.g., [2, 23]). This effect, as well as the length over which a cell can push neighbor cells away, has been previously described in cellular automata with one cell per lattice site [27-29].

Bipolar budding (Fig. 2B) is modeled as a combination of two sets of budding rules. The new bud will either form near the bud scar, in which case the same rules as those used for axial budding are used (Fig. 2B, Case 2), or the bud site will be on the end opposite to the previous bud scar, in which case polar budding rules are specified (Fig. 2B, Case 1). According to the polar budding rules, the cell can bud into the sites which are either 135° or 180° from the bud scar. If multiple sites are available, then one is randomly selected. If none of these sites are available, then one is randomly selected and the occupying cell is pushed out of the way, unless that area of the colony is considered too crowded, in which case the cell cannot bud. The polar and axial budding rules are used together to simulate the bipolar budding pattern of diploid cells. How often a cell buds at the distal pole, as opposed to the proximal pole, is affected by several factors [25]. For simplicity, in our model the bipolar budding pattern is set to follow axial budding rules half of the time and polar budding rules the other half of the time. The only exception to this rule is when a cell divides for the first time, in which case it will always bud according to the polar rules when budding in a bipolar pattern.

A parameter is set at the beginning of the simulation to indicate what fraction of cells will bud axially ($p_{axial}$). In a typical haploid colony 60-75% of cells will bud along an axial pattern and 20-40% of cells will bud in a bipolar pattern, while the remaining percentage of cells will bud randomly [30]. A $p_{axial}$ of 60% was used in our model for the haploid colonies. In a typical diploid colony, nearly all cells bud according to the bipolar pattern, with occasional random budding; in this case a $p_{axial}$ of 0% was used. Whenever a cell attempts to a bud, a random number between 0 and 1 is generated from a uniform distribution (Fig. 1). If this random number is less than $p_{axial}$, the cell will bud axially, otherwise the cell will bud in a bipolar fashion.

Nutrient concentrations were modelled using discrete nutrient packets. Three parameters related to nutrients are set at the beginning of the simulation. The first parameter is *nSteps*, the length of the random walk taken by a diffusing nutrient packet. The second parameter is *START_NUTRS*, the initial nutrient concentration, which sets the number of nutrient packets that all lattice sites will be given at the beginning of the simulation. The third parameter is *NUTRS_FOR_BUDDING*, the number of nutrient packets that a cell must consume before it can bud. At the beginning of each step, a random lattice site containing at least one nutrient packet is selected (Fig. 1). This nutrient packet takes a random walk of *nSteps* across the lattice. After a nutrient packet moves, a random cell is selected. If at least one nutrient packet is present at this cell's location, it will be consumed. If this cell has consumed *NUTRS_FOR_BUDDING* packets, it will attempt to bud according to the budding rules described above.

*Low-Nutrient Conditions*

Under low nitrogen conditions, diploid cells can switch from the bipolar budding pattern to a type of filamentous growth known a pseudohyphal growth [31]. Similarly, in low-nutrient conditions haploid cells can switch to an invasive filamentous growth [32-34]. In filamentous growth, new daughter cells are more elongated. These elongated cells will bud along the same direction and do not separate from one another, forming chains which resemble the hyphae seen in other fungi including pathogenic yeasts [35-37]. Like pseudohyphal growth, invasive growth in haploid cells is a result of cells budding along the same direction. In contrast to diploid cells, haploids extend their chains to grow downwards to penetrate agar medium [32, 34].

To model pseudohyphal growth in low-nutrient conditions, we incorporated a unipolar budding condition (Fig. 2C). The frequency of a cell budding according to this growth pattern is tied to the number of nutrient packets at the site occupied by a cell. Before any budding occurs, the probability that the selected cell will not bud according to the unipolar budding rules is calculated as the number of nutrient packets at that lattice site multiplied by a parameter $\Delta_{unipolar}$. This parameter is the reciprocal of 8 (the number of lattice sites surrounding a cell) times the number of nutrients a cell must consume to bud (*NUTRS_FOR_BUDDING*). Therefore, if there are $8 * \Delta_{unipolar}$ nutrient packets, then there are exactly enough nutrients for the cell to bud into all 8 lattice sites. This is the threshold for high-nutrient concentration. At this level of nutrients, the probability of budding in a normal budding pattern ($p_{normal}$) is 1. If there are fewer nutrients, then there is a chance that unipolar budding may occur. If nutrient levels rise above this threshold, the probability remains at 1. When a cell attempts to bud, a random number is generated from a uniform distribution between 0 and 1. If it is lower than $p_{normal}$, the cell will bud normally, and if it is higher than $p_{normal}$, it will grow according to the unipolar budding pattern. The lower the number of nutrient packets, the smaller the fraction and therefore the higher the chance of filamentous growth. If the initial nutrient concentration at every lattice site is sufficiently high, the simulation emulates the standard culture condition. When a cell buds according to the unipolar budding rules, the bud will occupy two lattice sites, one in front of the other along the mother-bud axis to create an elongated cell (Fig. 2C). The mother cell's bud scar will remain the same, rather than changing according to the direction of budding. This keeps the direction of future budding the same, simulating how cells bud in a chain in the same direction when undergoing filamentous growth. The lattice site representing the tip of the elongated cell has a bud scar set to be in the direction towards the mother cell, while the site closer to the mother cell has a bud scar of [0 0], such that it will not be able to bud. This occurs because this site represents the tail end of the elongated daughter cell, which is still attached to its mother cell, so nothing can bud from this end. Since filamentous growth primarily involves chains of cells that do not fully detach from each other, the unipolar budding condition cannot push other cells out of the way since cells do not bud in the middle of the chain. This limits the amount of unipolar budding that occurs in the middle of the colony, which represents actual colony growth more accurately; pseudohyphal growth in low-nutrient conditions occurs primarily at the colony edge to permit cells to search for nutrients [35, 36]. In the simulations, the parameter *UNIPOLAR_ON* controls whether cells will switch to unipolar budding in low nutrient conditions. If this parameter is false, cells will bud according to the normal budding rules regardless of the nutrient concentration.

*Magnetic Field Exposure*

There are yet to be experiments performed on the effects of MFs on yeast colony growth and morphology. Therefore, we assumed that the behaviour of a budding yeast colony under static MFs emerges from the division pattern of individual budding yeast cells that make up the colony. To implement this in our model, we incorporated the quantitative single-cell MF-budding angle data from Egami et al. [13]. Specifically, individual yeast cells were specified to bud within 30° to 150° of the direction of the applied homogeneous static MF. As this is a 2D model, the direction of the MF was constrained to the same plane as the yeast colony.

Whenever a cell buds in the presence of a MF, an MF bias condition is applied (Fig. 4). If a cell buds and pushes another cell out of the way, the direction in which that cell is pushed will also be biased in the direction of the MF vector ($\vec{B}$). To bias budding towards $\vec{B}$, first the array of possible lattice sites that the cell might bud into if no MF were present is determined. Next, the angle between each of these sites and the MF is calculated. The sites which are within the range of the angles set at the beginning of the simulation are chosen and the rest are removed as options. If there are no sites within that range of angles, then the list of possible sites determined at the start of the simulation remains unchanged. Since the model is based on a square lattice, which does not have eight-fold rotational symmetry, these MF biasing rules are not identical for every direction of the MF vector. When the MF vector is along one of the axes, we refer to the MF as an "axial" MF (Fig. 4A), whereas if the vector is along a diagonal, we refer to the MF as a "diagonal" MF (Fig. 4B).

The parameter *MF_STRENGTH* was introduced to control the strength of the MF. Each time the function for the MF bias is called (Fig. 1), a uniformly distributed random number between 0 and 1 is generated. If this number is less than the threshold set by *MF_STRENGTH*, the bias condition will be applied, otherwise it will not be applied. A strong MF is set when *MF_STRENGTH* is equal to 1, in which case the MF bias condition is applied every time. The model prioritizes budding into an empty site over budding along the direction of the MF, so in the case that the only unoccupied sites around the cell are outside of the space preferred by the MF (grey sites in Fig. 4), the cell will bud into those empty sites. An extra strong MF bias condition was created as well. The extra strong MF behaves similarly to the regular MF bias, except in the case when all the sites around the cell in the space preferred by the MF are occupied and there are unoccupied sites outside of this space. While in the normal MF bias condition, the cell would simply bud into one of those empty sites, the extra strong MF forces the cell to bud into a site in the preferred space and push the occupying cell out of the way. The extra strong MF condition can be applied by setting *MF_STRENGTH* to any number greater than 1.

**RESULTS AND DISCUSSION**

Here we use the simulation framework presented above to investigate the effects of nutrient conditions and magnetic fields on the growth and morphology of yeast colonies. Unless otherwise indicated, all results are for haploid cells without unipolar budding. The stopping condition was 320,000 timesteps for time-based simulations and 10,000 cells for cell count-based simulations. The MF bias parameter *MF_STRENGTH* was set to 1 for strong MFs and 0.5 for weak MFs. Cells were required to consume one nutrient packet to bud and nutrient diffusion was set as a 10-step random walk.

*Colony Growth*

To computationally investigate the effect of nutrient conditions and MFs on growth rate and budding angle of yeast cells, we determined the area, formation time, perimeter, and budding angle distribution of the colonies.

*Colony Area*

The area of the *in-silico* colonies was determined using MATLAB's *bwarea* function, giving roughly the number of pixels of the final colony image.

The final colony area decreased as the nutrient concentration decreased (Fig. 5). As the nutrient concentration decreases in our simulations, there are fewer nutrient packets diffusing on the lattice and therefore, on average, cells in the colony consume nutrients at a lower rate and bud less frequently. This result is in agreement with experimental results, where maximum colony area reached a larger size with increasing sugar concentrations [2].

The application of a strong axial MF decreased the final colony area independently of the nutrient concentration (Fig. 5). Weak axial and strong diagonal MFs caused a small decrease in the final colony area. As expected, similar results were obtained when the final number of cells was analyzed instead of the final colony area (Fig. S1).

Though strong axial MFs and low-nutrient conditions both reduced colony growth, the nutrient condition is predicted to have a greater effect on the growth rate of the colony compared to the application of MFs (Fig. 5; Fig. S1). Overall, these results suggest that nutrient limitation in the microenvironment and MF exposure could mitigate growth in yeast infections and colony biofilms.

*Colony Formation Time*

Colony formation time was calculated in the simulations as the time necessary for colonies to reach 10,000 cells.

As the nutrient concentration decreased, colony formation time increased (Fig. 6).

The effect of magnetic fields on colony size followed a consistent pattern across the nutrient conditions (Fig. 6). Axial MFs and strong diagonal MFs increased the number of timesteps required to reach the final colony size.

The colony formation time results presented in Figure 6 correspond inversely to the colony area results in Figure 5, meaning that a larger final colony area for a fixed number of timesteps and smaller number timesteps to reach a given colony size both represent higher growth rates.

*Colony Perimeter*

The perimeter of the colony was determined from a list of coordinates of the boundary pixels; a sum of the distance from each pixel to the next on the colony edge was obtained from these coordinates.

Colony perimeter decreased with nutrient concentration (Fig. 7). These results are consistent with a decrease in colony area (Fig. 5) and an increase in colony formation time (Fig. 6) in the low-nutrient condition. The nutrient-dependent trends observed in colony perimeter can be attributed to the decreased growth rate of the colony at low-nutrient concentrations, like the decreased growth rate effect on colony area and colony formation time.

MFs had little effect on the colony perimeter in the rich-nutrient and low-nutrient conditions (Fig. 7). The lack of an effect of MFs on colony perimeter can be attributed to the fact that the perimeter is also influenced by irregularities at the colony boundary. Specifically, if the boundary fluctuations were not a factor, a decrease in colony area in the presence of a MF (Fig. 5) would lead to a decrease in colony perimeter. On the other hand, if colony area was not a factor, the increase in boundary fluctuations when exposed to MFs (see *Boundary Fluctuations* subsection and Fig. 13), would cause an increase in colony area. Since both colony area and boundary fluctuation effects are present, these opposing influences on perimeter cancel each other out (Fig. 7). Boundary irregularities are more accurately determined from the boundary fluctuations measurement described in the see *Boundary Fluctuations* subsection that corrects for perimeter variation due to various factors.

*Budding Angle*

The MF bias rules were based on experimental data that showed that the angle between the mother-bud axis and the direction of the MF was most often between 30° and 150° [13]. We calculated this angle each time a cell budded over the course of our simulations. Regardless of ploidy or the nutrient condition, the frequencies of each budding angle for haploid colonies were about equal when no MF was present (Fig. 8; Fig. S7), in agreement with experimental results [13]. When an MF was applied, the number of times the budding angle was between 30° and 150° was greater than outside of this range (Fig. 8; Fig. S7), in agreement with experimental results [13]. This effect was more prominent with strong MFs as opposed to weak MFs in haploid (Fig. 8) and diploid colonies (Fig S7). The budding angle distributions were similar for axial MFs and diagonal MFs (data not shown).

## *Colony Morphology*

To quantify the effect of nutrient conditions and MF exposure on colony morphology, we investigated the roundness, elongation, solidity, and boundary fluctuations of the *in-silico* yeast colonies.

*Roundness*

Roundness is the ratio of the area of the colony to the area of a circle having the same perimeter as the "convex hull" of the colony:

$$Roundness = \frac{4\pi A}{P_C^2} \qquad (1)$$

where $A$ is the area of the colony and $P_C$ is the perimeter of the convex hull. The convex hull is the smallest convex shape which contains the entire colony, where "convex" implies that there are no corners that bend inward. This gives a measurement of how much the colony shape differs from a circle, without much sensitivity to smaller fluctuations along the boundary. As roundness decreases it indicates a less circular colony. A colony with an irregular boundary can still yield a high roundness measure if its general shape is circular. Note that the inverse Eqn. (1) yields a relationship that similar to the scale-invariant "P2A" ratio, which has also been used to describe the deviation of colony shape from a circle [2].

Low-nutrient conditions decrease the roundness of the colonies (Fig. 9). This indicates that the colonies become less circular as nutrient concentrations decrease. This occurs because yeast cells regularly fail to bud in low-nutrient concentrations, resulting in more unoccupied grid lattice sites and a more irregular colony. The decrease in roundness may also be related to petal formation on the colony boundary, which occurs due to increased intercellular competition over low numbers of diffusing nutrients at the colony rim [2].

Strong axial magnetic fields decreased the roundness of the colonies independently of the nutrient concentration (Fig. 9). Weak axial MFs had a similar effect, but to a lesser degree. Diagonal MFs did not affect the roundness of the colonies.

*Elongation*

Elongation is the ratio of the width to the height of the bounding box of an *in-silico* colony; a higher ratio increases indicates a more elongated colony. The bounding box of the colony image is the smallest possible rectangle which contains the entire area of the colony. This was determined by finding the angle of orientation of the colony with the orientation property of the *regionprops* function in MATLAB and then using this to orient the colony along the y-axis, and finally by applying *regionprops* to find the bounding box property. It was necessary to orientate the colony along the y-axis as the bounding box property only finds the smallest possible rectangle which is not rotated. By rotating the colony, colonies that were stretched in any direction could be compared without the direction of elongation influencing the results.

Nutrient concentration had no effect on colony elongation (Fig. 10).

The application of axial MFs greatly increased the elongation of the colonies, with strong axial MFs having an even greater effect than weak axial MFs (Fig. 10). Therefore, the exposure of budding yeast cells to axial magnetic fields is predicted to generate less circular colonies, in agreement with colony roundness results (Fig. 9). Diagonal MFs had no effect on the elongation of colonies (Fig. 10), also in agreement with colony roundness results (Fig. 9). However, diagonal MFs did begin to have an effect in low nutrient conditions, increasing the elongation of colonies, when the strength of the MF was increased to the extra strong condition (Fig. 11). Extra strong axial MFs also had an increased effect compared to strong axial MFs. Overall, these results indicate that MFs could provide a novel way to control the morphology of cells growing on biomaterials.

*Solidity*

Solidity is the ratio of the area of the colony to the area of the convex hull, which measures the cellular density of the colony:

$$Solidity = \frac{A}{A_c} \qquad (2)$$

where $A$ is the area of the colony and $A_c$ is the area of the convex hull.

The solidity of the colonies was found to decrease as the nutrient concentration decreased (Fig. 11). This decrease in solidity can be attributed to the fact that the ability of cells to bud and fill up empty lattice sites in our simulations decreases along with the nutrient concentration.

Axial MFs decreased the solidity of colonies slightly, with strong axial MFs having a more pronounced effect than weak axial MFs (Fig. 12). Diagonal MFs had no effect on colony solidity.

*Boundary Fluctuations*

The boundary fluctuation measure captures irregularities in the colony shape, such as petals and asymmetries that form at the colony rim. Specifically, the boundary fluctuation is the ratio of the standard deviation of the normalized radial lengths ($\sigma_{RL}$) to the mean normalized radial lengths ($\mu_{RL}$):

$$Boundary\ Fluctuation = \frac{\sigma_{RL}}{\mu_{RL}}. \qquad (3)$$

The radial length is the distance from the center of the colony to a point on the boundary; the centroid was calculated using the *regionprops* function in MATLAB. Each radial length is normalized by dividing it by the maximum radial length.

Figure 13 shows that boundary fluctuations increase as the nutrient concentration decreases. This increase in irregularity at the colony rim in the low-nutrient condition is in agreement with experimental results [2].

Axial MFs caused a large increase in the boundary fluctuations, with strong axial MFs causing an even larger increase than weak axial MFs. Diagonal MFs did not affect boundary fluctuations, but like with elongation, extra strong diagonal MFs increased boundary fluctuations, particularly in the low-nutrient condition (Fig. 14).

*Effects of Axial Versus Diagonal Magnetic Fields on Colony Growth and Morphology*

To investigate why axial MFs had a greater overall effect than diagonal MFs on the growth and morphology of haploid yeast colonies, we performed the corresponding simulations on diploid yeast colonies that follow a different budding pattern. For a diploid colony, the normal budding pattern is bipolar (Fig. 2B) when nutrients are readily available but can switch to pseudohyphal growth (Fig. 2C) when nutrients are scarce.

In the absence of MFs and pseudohyphal growth, diploid colonies reached a larger final area than haploid colonies (Fig. 15; Fig. S2); this effect was also reflected in the shorter times for diploid colonies to reach 10,000 cells (Fig. S3) and in larger diploid colony perimeters (Fig. S4), when compared to haploid colonies (Figs. 6-7). We attribute the larger size of the diploid colonies compared to haploid colonies in our simulations to differences in their respective budding patterns. Diploid cells may have a higher success rate for budding because of less crowding in the surrounding lattice sites, resulting from budding switching between the different poles of diploid cells.

For haploid colonies undergoing pseudohyphal growth, axial MFs and diagonal MFs had a substantial effect on colony morphology, resulting in horizontally or diagonally stretched colonies, respectively, particularly in the low-nutrient condition (Fig. 3).

Diploid colonies that could switch to pseudohyphal growth had a lower cell count at the end of the simulations in the low-nutrient condition (on average 7,500 cells in the low-nutrient condition compared to on average 11,500 cells in the nutrient-rich condition) in agreement with the nutrient-dependent growth results for haploid yeast colonies (Figs. 3 & 5-6).

The density of haploid colonies (Figs. 3 & 12) and diploid colonies (Fig. S6) were less dense under nutrient limitation and MF exposure as result of these colonies containing more holes (i.e., unoccupied lattice sites). This nutrient- and MF-dependent decrease in density could in principle reduce antimicrobial resistance in colony biofilms by enhancing the penetration of antimicrobial drugs into the colony biofilm.

Diagonal MFs resulted in more elongated diploid colonies when there was no pseudohyphal growth (Fig. 16) compared to haploid colonies whose elongation was not influenced by weak or strong diagonal MFs (Fig. 10). Diagonal MFs also increased the boundary fluctuations in diploid colonies (Fig. 17) compared to haploid colonies, which were not influenced by weak or strong diagonal MFs (Fig. 13).

Overall, these results indicate that the interaction between MFs and ploidy-dependent budding patterns determines if an applied axial MF or diagonal MF will yield the greatest effect on colony growth and morphology.

**CONCLUSIONS**

We developed a 2D lattice-based simulation framework to spatiotemporally investigate the interplay between biological phenomenon and physical forces in yeast. This novel computational framework was able to reproduce experimental results obtained under nutrient-varying conditions [2]. The simulated budding angle distributions were also found to be in agreement experimental budding angle distributions under magnetic field exposure [13]. The lattice-based framework was then used to make predictions on the effects of magnetic fields on colony growth and morphology.

Simulations of the lattice-based model accurately captured the colony area and colony formation time in agreement with previous experiments on budding yeast [2]. This was achieved by modeling the diffusion of nutrient molecules that cells on the lattice consumed to divide. The nutrient packets could be replaced with drug molecules to allow drug delivery and the formation of drug-resistant biofilms. We anticipate that this framework will also be suitable for modelling clonal interference among drug resistance mutants [38] and that it could be expanded to three dimensions to model the invasive growth (i.e., the vertical penetration of the filaments into the agar or tissue) of haploid yeast cells exposed to nutrient-limiting conditions [32-34], as well as more intricate pattern formation in eukaryotic biofilms [2]. Augmenting the framework to include more complex yeast behaviors, such as mating between haploid cells, and meiosis and sporulation in diploid cells facing extreme nutrient-depletion conditions [39-41] will be important areas to explore in future work.

The lattice-based framework was used to generated novel predictions on how magnetic fields affect the spatiotemporal characteristics of yeast colonies by incorporating experimental data on the budding angle of individual yeast cells exposed to magnetic fields [13]. Strong magnetic fields were shown in our simulations to exhibit wide-ranging effects on yeast colony growth and morphology in low-nutrient and high-nutrient conditions. The decrease in colony growth rate due to applied magnetic fields opens the possibility of using electromagnetic fields to control fungal infections and mitigate biofilm formation. Similarly, the decrease in colony density under magnetic field exposure may increase the ability of antimicrobial drugs to penetrate biofilms when they are exposed to electromagnetic fields. Finally, the elongation of the colonies along the magnetic field vector in our simulations may provide a novel way to control to guide cellular growth on biomaterials.

Comparison between haploid and diploid colony simulations revealed that the interaction between the magnetic field and the budding pattern determines if an applied axial or diagonal magnetic field will yield the greatest effect on colony growth and morphology. The differences between the implementation of the axial and diagonal magnetic fields in our simulations are a limitation of using a square lattice and not a true electromagnetic phenomenon. This may be resolved in future work by using a hexagonal mesh rather than a square lattice [42-44]. Electromagnetic fields have also been observed to effect yeast cell proliferation in a frequency-dependent manner [12], which could be simulated using the lattice-based framework by incorporating alternating magnetic fields. The mechanism by which budding yeast cells orient themselves to magnetic fields is an open question, though it has been hypothesized to be a result of polarized microtubules [13]. Further experiments on yeast cells are required to validate our magnetic field simulation results and to elucidate the mechanisms underlying these biomagnetic effects.

Quantitative spatiotemporal modelling, such as the lattice-based framework presented in this study, combined with the development of electromagnetic devices to perform controlled laboratory experiments will provide powerful tools for elucidating bioelectromagnetic mechanisms in microbial communities. Uncovering such mechanisms may one day allow electromagnetic fields to be used to control growth and biofilm formation in yeast and other microorganisms including magnetotactic bacteria [45], as well as to enhance the efficacy of antimicrobial drugs.

**Author Contributions**

DC conceptualized and supervised the study. RH and DC developed the lattice-based modeling framework. RH developed the MATLAB code and performed the simulations. RH and DC analyzed the results. DC and RH wrote the manuscript.

**Funding**

DC was supported by funding from an NSERC Discovery Grant (RGPIN-2020-04007), an NSERC Discovery Launch Supplement (DGECR-2020-00197), and the University of Alberta. RH was supported by a 2020 Alberta Innovates Summer Research Studentship and a 2021 NSERC Undergraduate Student Research Award.

**Conflict of Interest**

The authors declare that the research was conducted in the absence of any commercial or financial relationships that could be construed as a potential conflict of interest.

**Acknowledgments**

The authors would like to thank Akila Bandara, Lesia Guinn, and Dr. Jay Newby for helpful comments on the manuscript.

# FIGURES

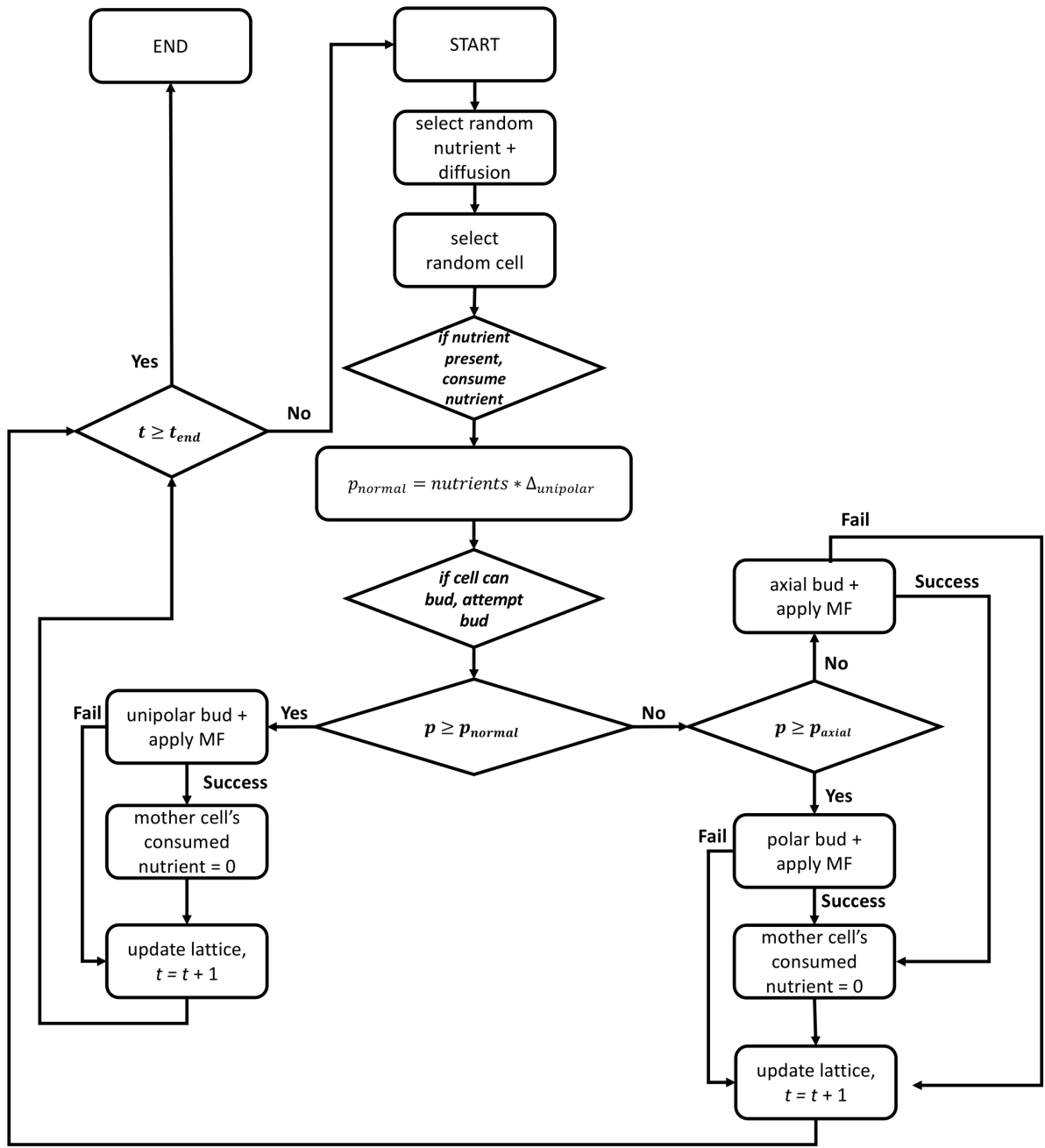

**Figure 1. Flow diagram of the lattice algorithm for simulating the effects of nutrient concentration and magnetic field (MF) exposure on yeast colonies.** Rectangles represent a step in the program while diamonds represent an evaluation made of the state of the lattice site to determine the next step to take. $p_{normal}$ is the probability of the cell budding into an axial or bipolar pattern, rather than the unipolar pattern occurring in low-nutrient conditions. $p_{axial}$ is the probability of the cell budding in an axial pattern. $\Delta_{unipolar}$ is the fixed amount by which $p_{normal}$ increases with each nutrient packet consumed. The current timestep is represented by $t$ and $t_{end}$ is the total number of timesteps for which the algorithm is set to run.

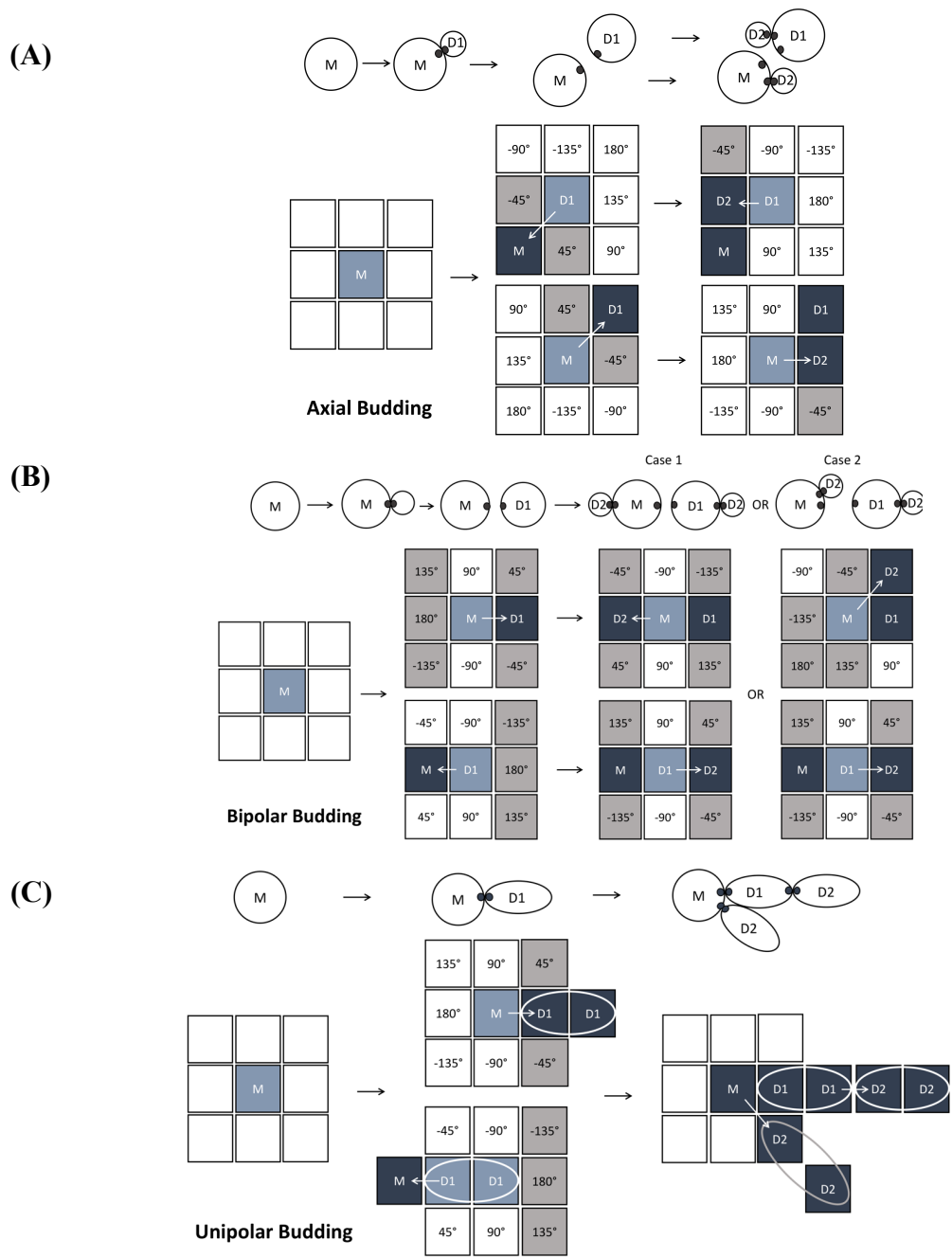

**Figure 2. Implementation of yeast budding patterns in the 2D-lattice model.** Diagrams of the axial budding pattern in haploid cells (A), the bipolar budding pattern in diploid cells (B), and the unipolar budding pattern in nutrient limiting conditions (C) (see main text for details). In each of these diagrams, *M* denotes the original mother cell and *D1* and *D2* denote the second and third generation of buds, respectively. A black circle on the cell schematics indicates a bud scar (or a birth scar) on the cell. On the grids, a white arrow indicates the direction of the dividing cell's bud scar, a white square an unoccupied lattice site, a light blue square a lattice site occupied by a dividing cell, a dark blue square a lattice site occupied by cells that are not dividing in the current timestep, and a grey square a possible lattice site into which a dividing cell can bud.

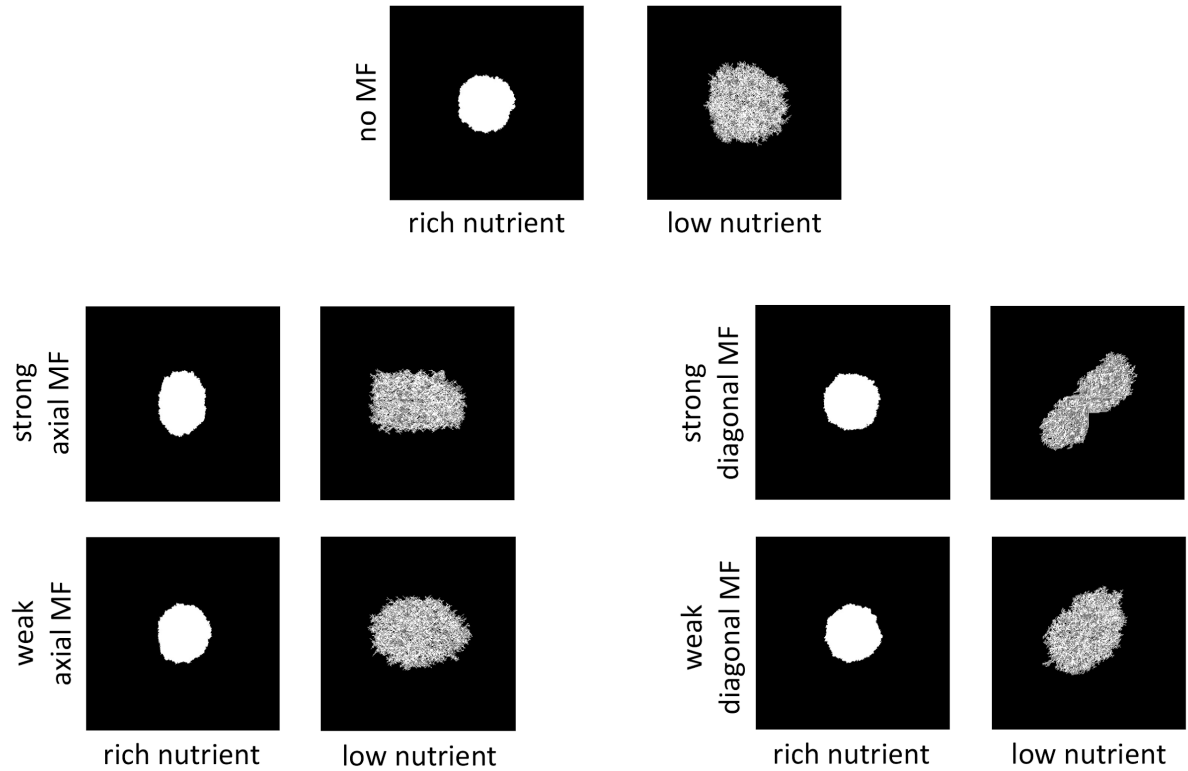

**Figure 3. Simulated haploid colonies with filamentous growth under different nutrient and magnetic field (MF) conditions.** Parameters were set as follows: rich-nutrient condition: *START_NUTRS* = 20, low-nutrient condition: *START_NUTRS* = 2; *nSteps* = 10; $p_{axial}$ = 0; diagonal MF direction: *MAGNETIC_FIELD* = [1 1], axial MF direction: *MAGNETIC_FIELD* = [1 0]; no MFs: *MF_STRENGTH* = 0, weak MFs: *MF_STRENGTH* = 0.5, strong MFs: *MF_STRENGTH* = 1; *UNIPOLAR_ON* = true.

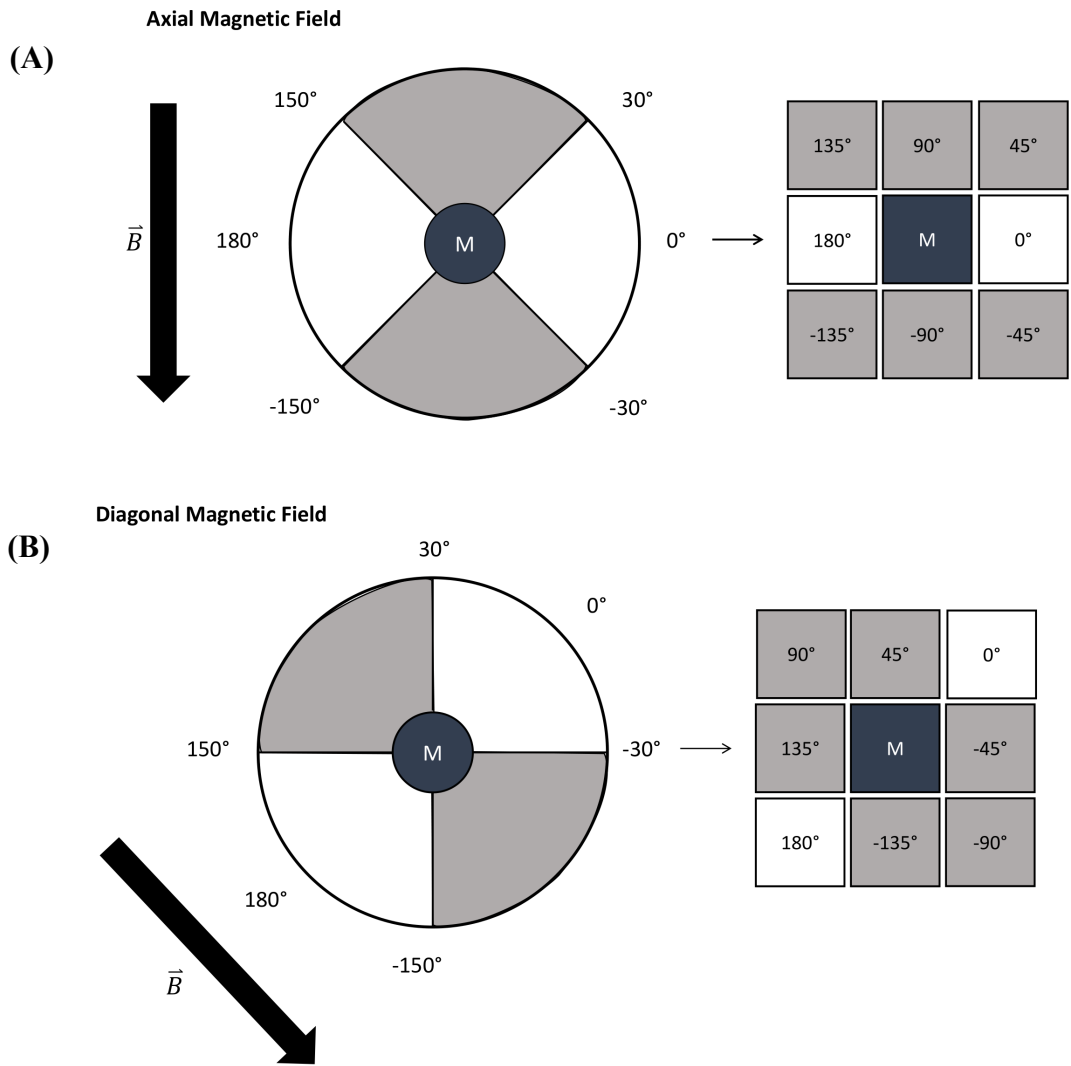

**Figure 4. Diagram of the implementation of magnetic fields in the 2D-lattice model.** Angles are measured with respect to the line perpendicular to the magnetic field ($\vec{B}$). **(A)** A magnetic field applied along the y-axis biased budding in the 30º to 150º or -30º to -150º range, which is shown in grey on the schematic (left panel). Translating this to a grid biases cells towards budding into the lattice sites in the top and bottom rows shown in grey (right panel). Similarly, a magnetic field applied along the x-axis would bias cells towards budding into lattice sites in the left and right columns (not shown). **(B)** A magnetic field applied along the diagonal biased budding in the 30º to 150º or -30º to -150º range, which is shown in grey on the schematic (left panel). Translating this to a grid biases cells towards budding into the corner lattice, shown in grey (right panel).

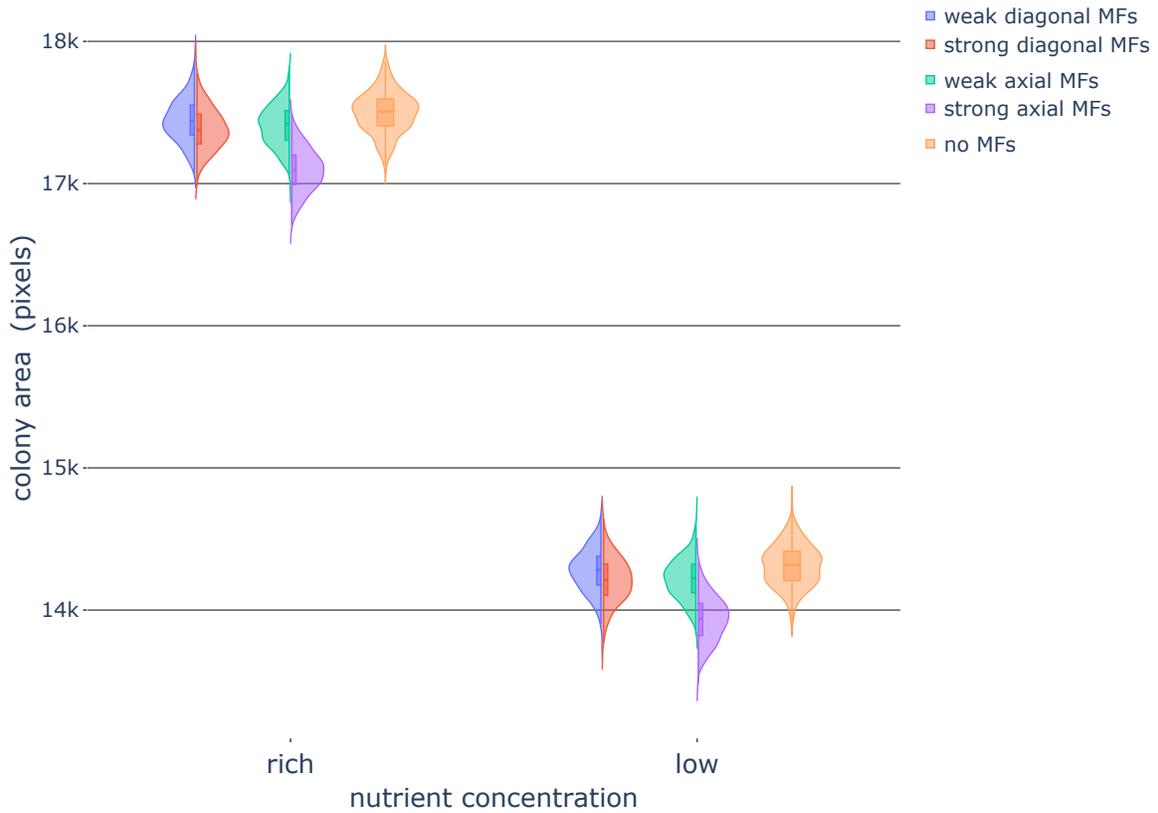

**Figure 5. Colony area under different nutrient and magnetic field conditions.** Violin plots of yeast colony area after 320,000 timesteps, when influenced by various magnetic field (MF) directions and strengths under rich-nutrient and low-nutrient conditions. Box plots within the violin plots denote the mean, quartiles, and outliers. Parameters were set as follows: rich-nutrient condition: *START_NUTRS* = 20, low-nutrient condition: *START_NUTRS* = 2; *nSteps* = 10; $p_{axial}$ = 0.6; no MFs: *MF_STRENGTH* = 0, weak MFs: *MF_STRENGTH* = 0.5, strong MFs: *MF_STRENGTH* = 1.

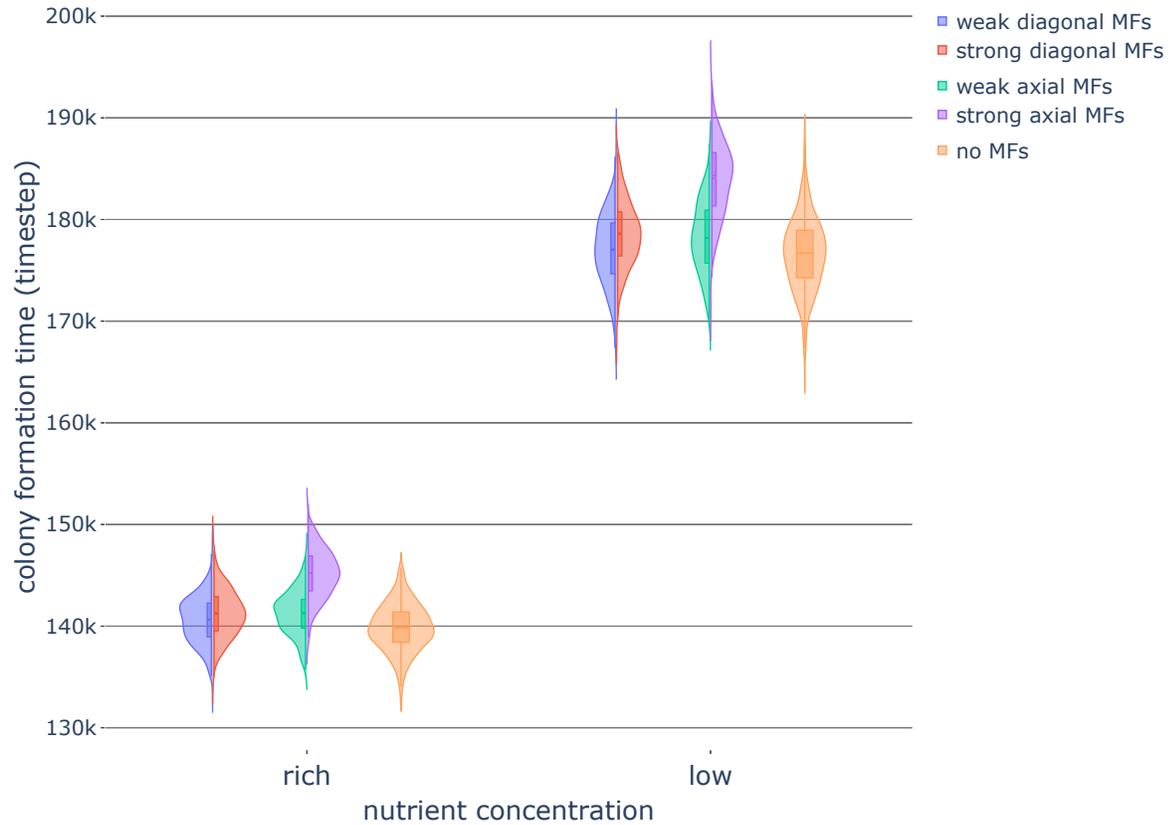

**Figure 6. Colony formation time under different nutrient and magnetic field conditions.** Violin plots of the time to form a colony of 10,000 cells (starting from a single cell), when influenced by various magnetic field (MF) directions and strengths under low-nutrient and high-nutrient conditions. Box plots within the violin plots denote the median, quartiles, and outliers. Parameters were set as follows: rich-nutrient condition: *START_NUTRS* = 20, low-nutrient condition: *START_NUTRS* = 2; *nSteps* = 10; $p_{axial}$ = 0.6; no MFs: *MF_STRENGTH* = 0, weak MFs: *MF_STRENGTH* = 0.5, strong MFs: *MF_STRENGTH* = 1.

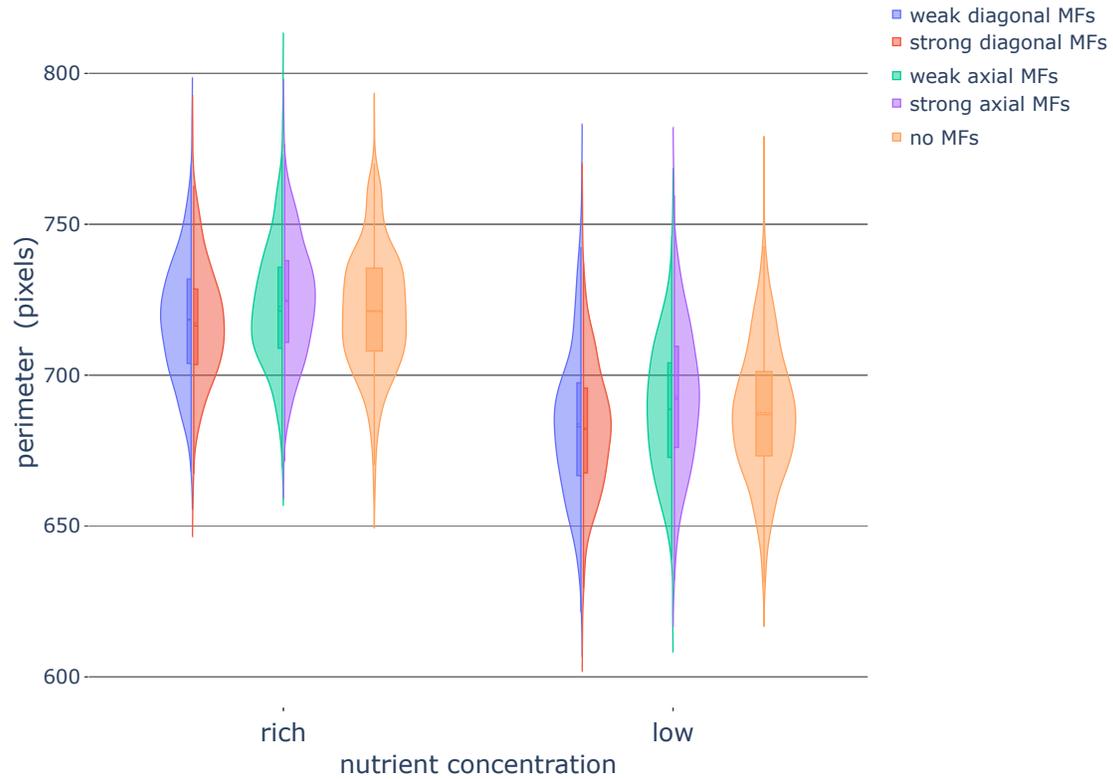

**Figure 7. Colony perimeter under different nutrient and magnetic field conditions.** Violin plot of colony perimeters of colonies in various nutrient conditions, when exposed to magnetic fields (MFs) of different strengths and directions. Box plots within the violin plots denote the median, quartiles, and outliers. Parameters were set as follows: rich-nutrient condition: *START_NUTRS* = 20, low-nutrient condition: *START_NUTRS* = 2; *nSteps* = 10; $p_{axial}$ = 0.6; no MFs: *MF_STRENGTH* = 0, weak MFs: *MF_STRENGTH* = 0.5, strong MFs: *MF_STRENGTH* = 1.

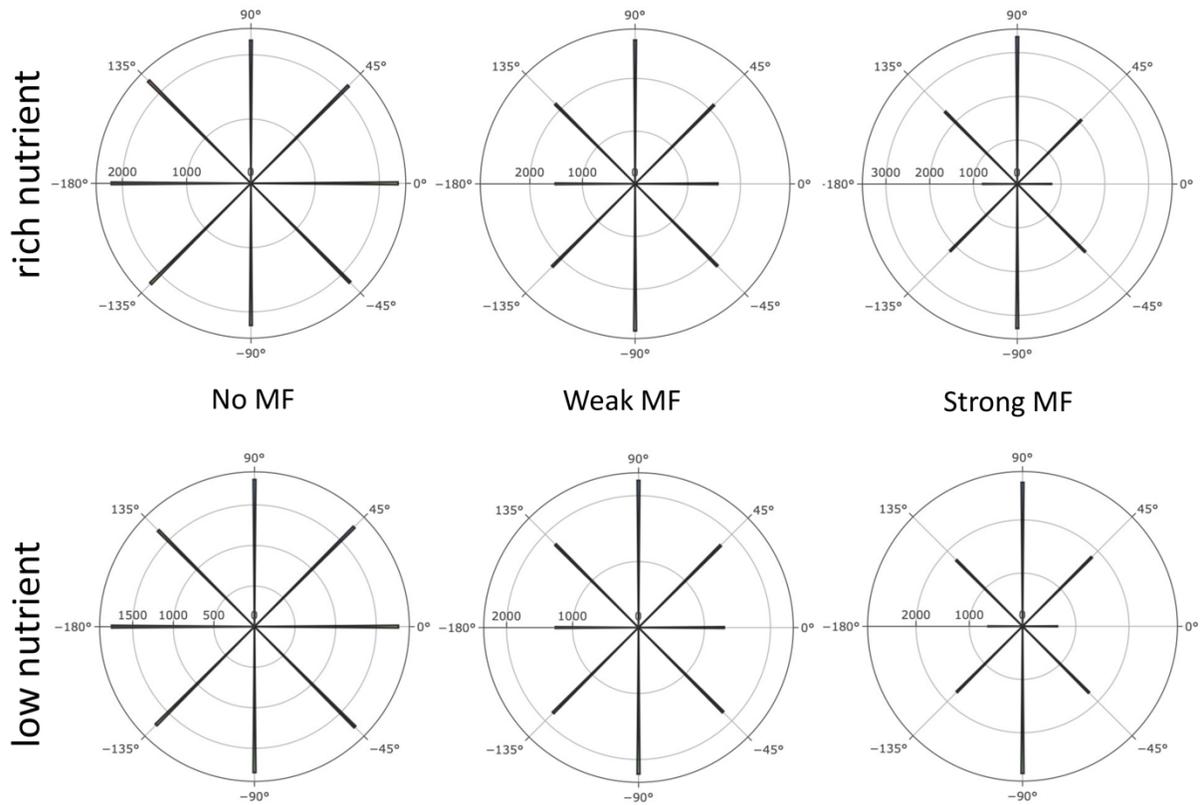

**Figure 8. Budding angle distributions for haploid colonies under different nutrient conditions of magnetic field strengths.** Plot of the frequency of the angle between the magnetic field direction and the mother-bud axis. Parameters were set as follows: rich-nutrient condition: *START_NUTRS* = 20, low-nutrient condition: *START_NUTRS* = 2; *nSteps* = 10; $p_{axial}$ = 0.6; *MAGNETIC_FIELD* = [1 0]; no MFs: *MF_STRENGTH* = 0, weak MFs: *MF_STRENGTH* = 0.5, strong MFs: *MF_STRENGTH* = 1; *UNIPOLAR_ON* = false.

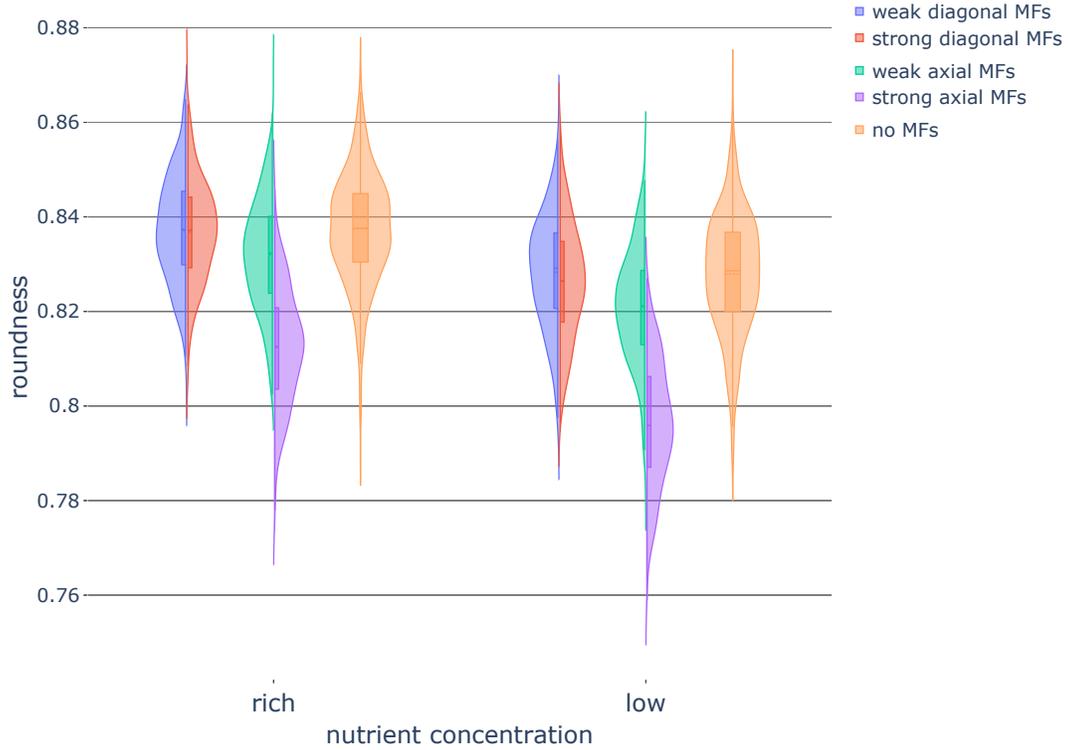

**Figure 9. Colony roundness under different nutrient and magnetic field conditions.** Violin plots of colony roundness under different nutrient conditions and different magnetic field (MF) directions and strengths. Box plots within the violin plots denote the median, quartiles, and outliers. Parameters were set as follows: rich-nutrient condition: *START_NUTRS* = 20, low-nutrient condition: *START_NUTRS* = 2; *nSteps* = 10; $p_{axial}$ = 0.6; no MFs: *MF_STRENGTH* = 0, weak MFs: *MF_STRENGTH* = 0.5, strong MFs: *MF_STRENGTH* = 1.

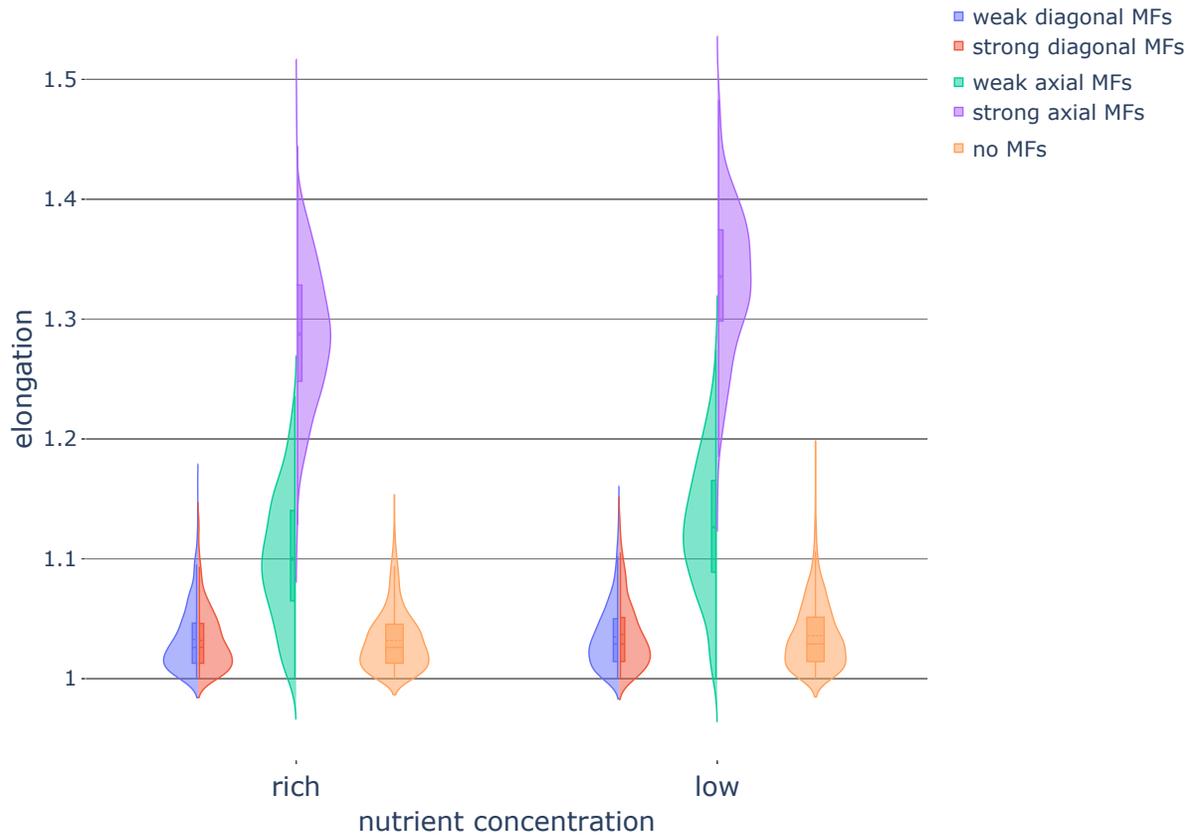

**Figure 10. Colony elongation under different nutrient conditions and applied magnetic fields.** Violin plot of colony elongation under various nutrient concentrations and exposure to magnetic fields (MFs) of different strengths and directions. Box plots within the violin plots denote the median, quartiles, and outliers. Parameters were set as follows: rich-nutrient condition: *START_NUTRS* = 20, low-nutrient condition: *START_NUTRS* = 2; *nSteps* = 10; $p_{axial} = 0.6$; no MFs: *MF_STRENGTH* = 0, weak MFs: *MF_STRENGTH* = 0.5, strong MFs: *MF_STRENGTH* = 1.

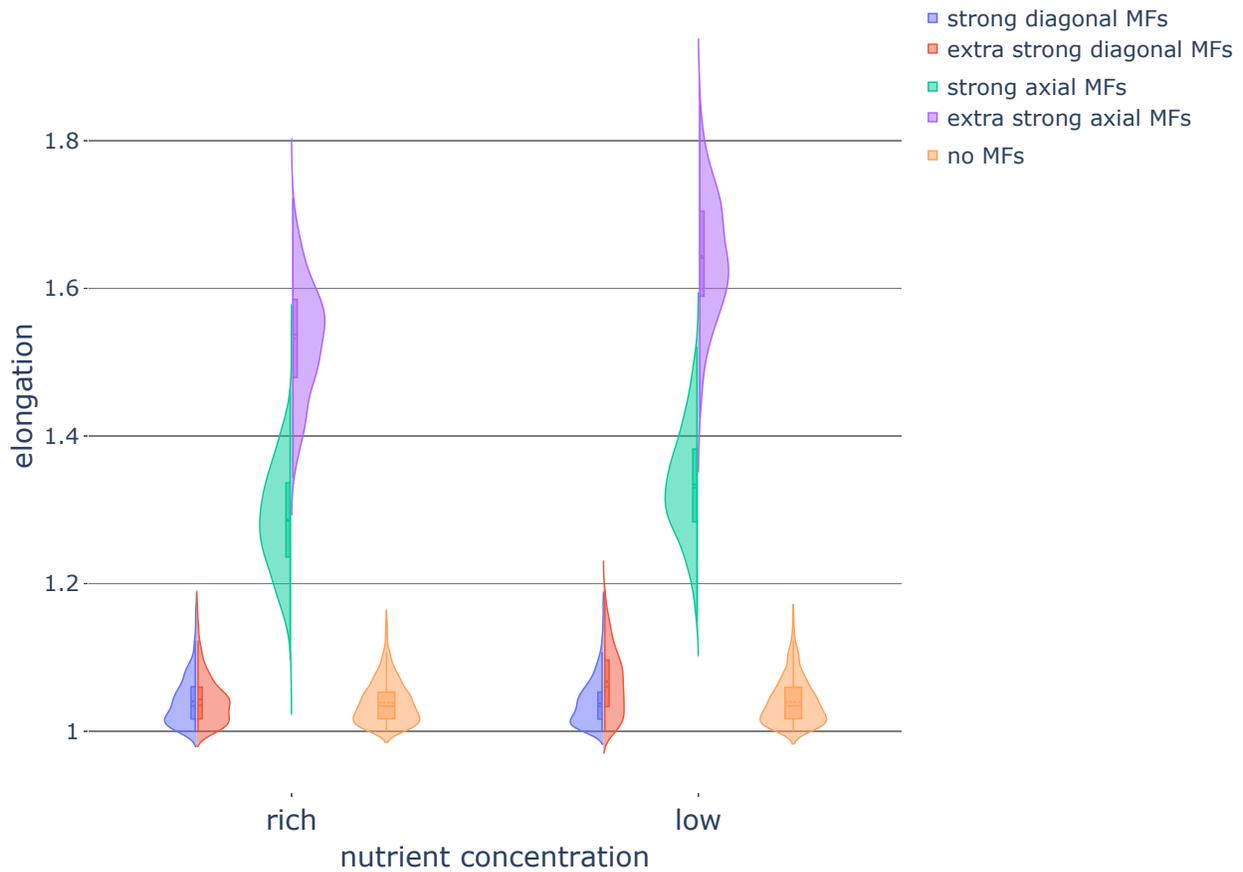

**Figure 11. Colony elongation under different nutrient conditions and strong and extra strong applied magnetic fields.** Violin plot of colony elongation under various nutrient concentrations and exposure to strong and extra strong magnetic fields (MFs) of different directions. Box plots within the violin plots denote the median, quartiles, and outliers. Parameters were set as follows: rich-nutrient condition: *START_NUTRS* = 20, low-nutrient condition: *START_NUTRS* = 2; *nSteps* = 10 steps; $p_{axial} = 0.6$; no MFs: *MF_STRENGTH* = 0, strong MFs: *MF_STRENGTH* = 1, extra strong MFs: *MF_STRENGTH* = 2.

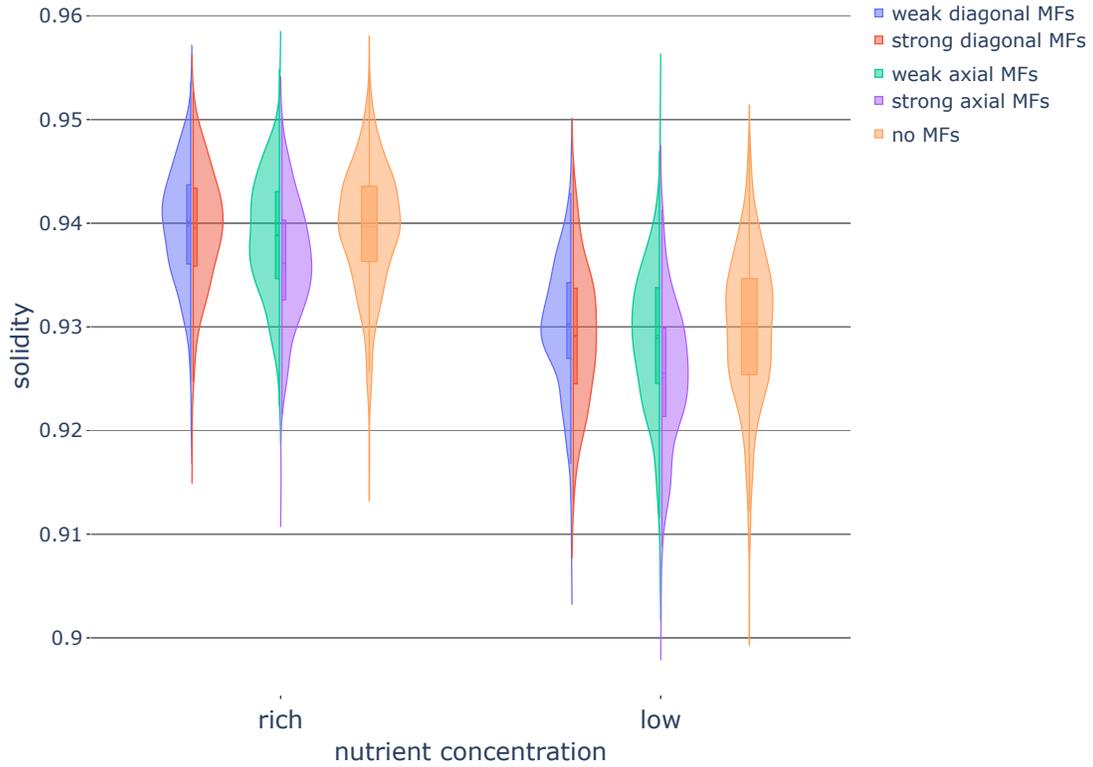

**Figure 12. Colony solidity under different nutrient and magnetic field conditions.** Violin plot of colony solidity in various nutrient conditions while exposed to magnetic fields (MFs) of different strengths and directions. Box plots within the violin plots denote the median, quartiles, and outliers. Parameters were set as follows: rich-nutrient condition: *START_NUTRS* = 20, low-nutrient condition: *START_NUTRS* = 2; *nSteps* = 10; $p_{axial}$ = 0.6; no MFs: *MF_STRENGTH* = 0, weak MFs: *MF_STRENGTH* = 0.5, strong MFs: *MF_STRENGTH* = 1.

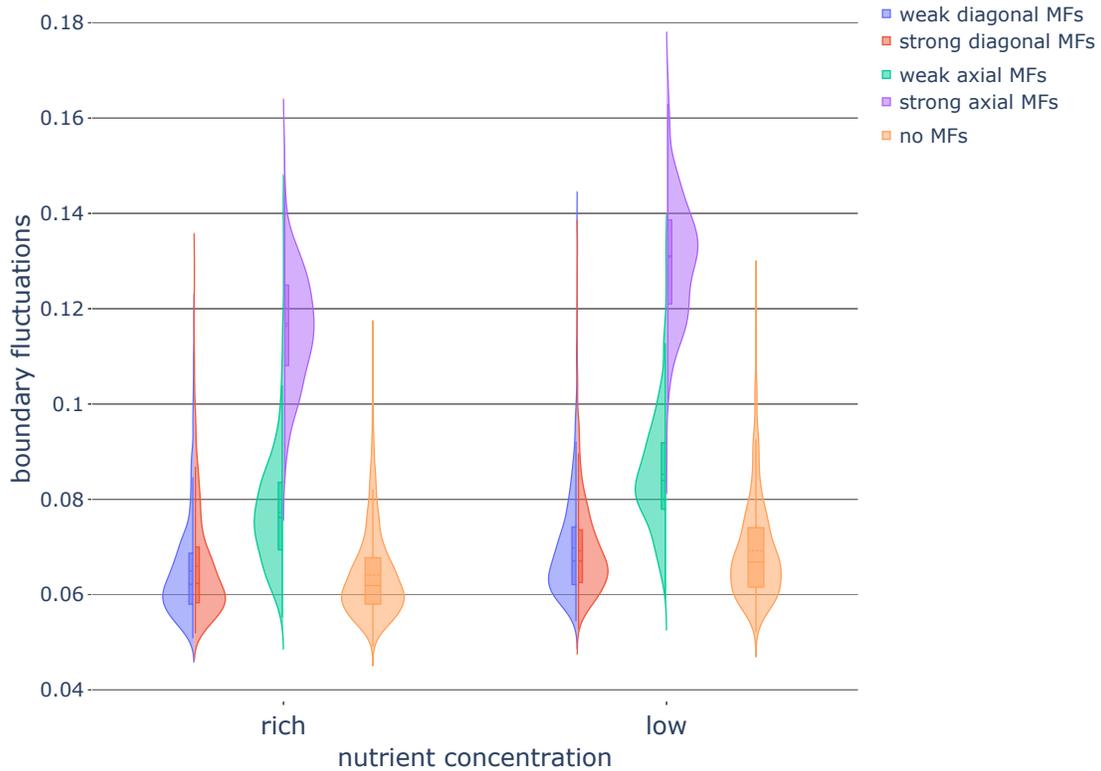

**Figure 13. Colony boundary fluctuations under different nutrient and magnetic field conditions.** Violin plots of elongation of colonies in various nutrient conditions and exposed to magnetic fields (MFs) of different strengths and directions. Box plots within the violin plots denote the median, quartiles, and outliers. Parameters were set as follows: rich-nutrient condition: *START_NUTRS* = 20, low-nutrient condition: *START_NUTRS* = 2; *nSteps* = 10; $p_{axial} = 0.6$; no MFs: *MF_STRENGTH* = 0, weak MFs: *MF_STRENGTH* = 0.5, strong MFs: *MF_STRENGTH* = 1.

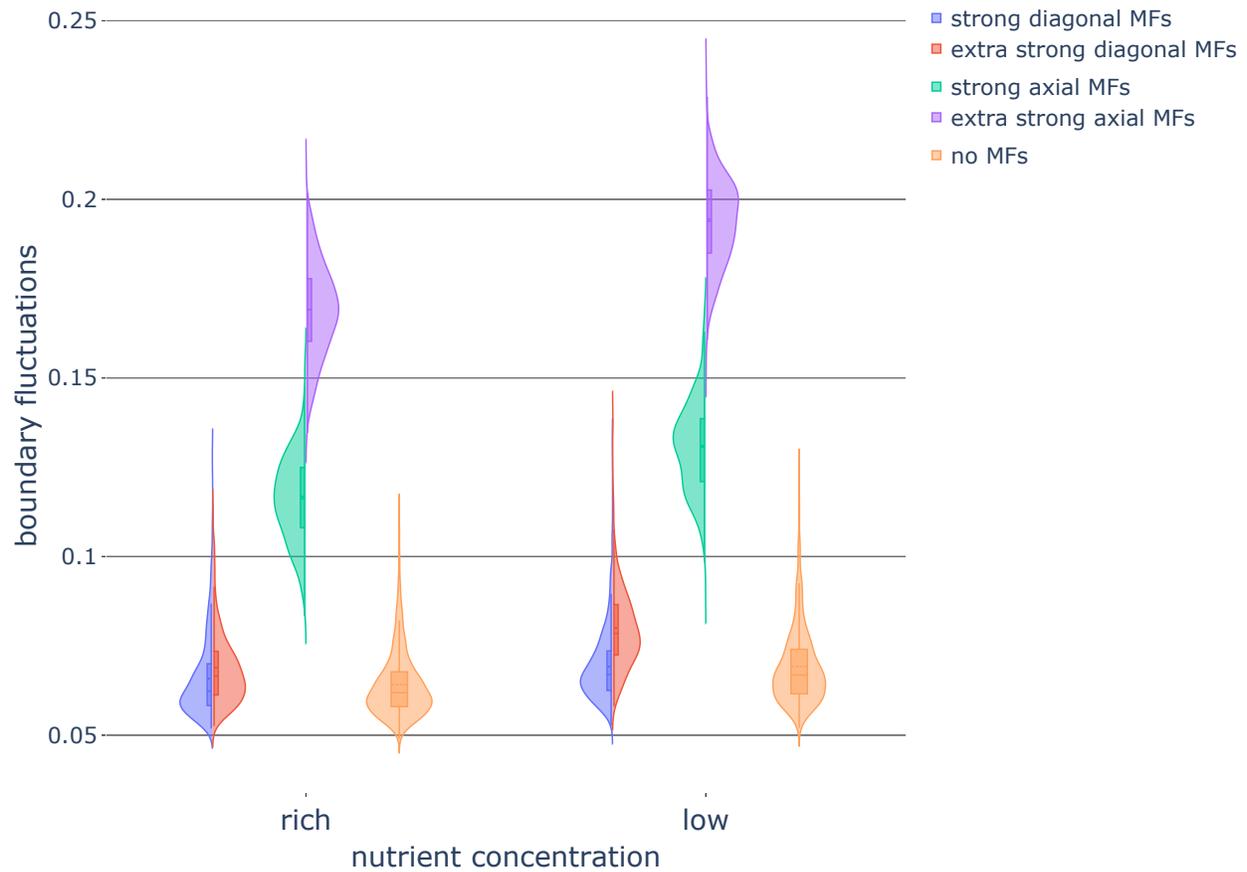

**Figure 14. Colony boundary fluctuations under different nutrient and strong and extra strong magnetic field conditions.** Violin plots of elongation of colonies in various nutrient conditions and exposed to strong and extra strong magnetic fields (MFs) of different directions. Box plots within the violin plots denote the median, quartiles, and outliers. Parameters were set as follows: rich-nutrient condition: *START_NUTRS* = 20, low-nutrient condition: *START_NUTRS* = 2; *nSteps* = 10; $p_{axial}$ = 0.6; no MFs: *MF_STRENGTH* = 0, strong MFs: *MF_STRENGTH* = 1, extra strong MFs: *MF_STRENGTH* = 2.

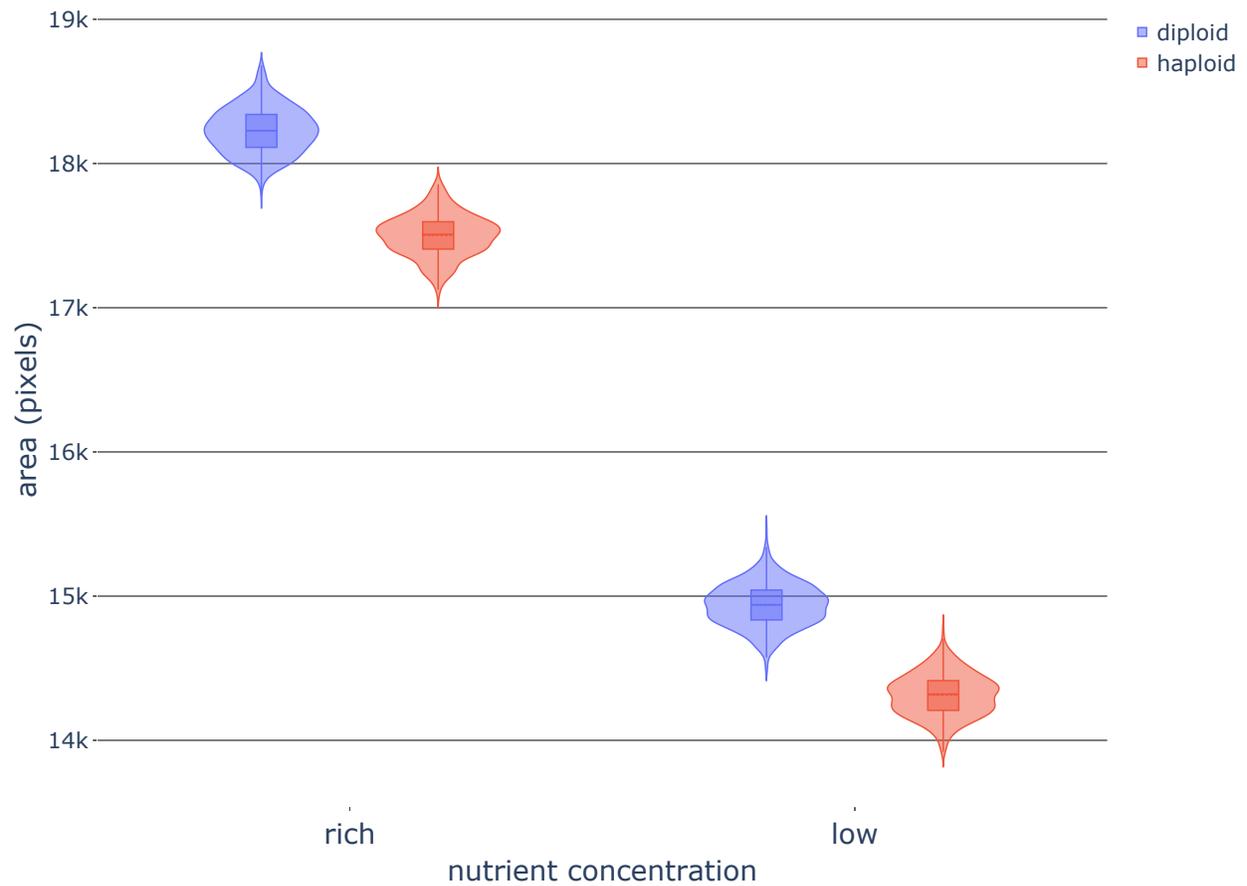

**Figure 15. Colony area of diploid and haploid colonies without pseudohyphal growth under different nutrient concentrations.** Violin plots of yeast colony area after 320,000 timesteps for different ploidies under rich-nutrient and low-nutrient concentrations with diffusion and no magnetic fields (MFs). Box plots within the violin plots denote the mean, quartiles, and outliers. Parameters were set as follows: rich-nutrient condition: *START_NUTRS* = 20, low-nutrient condition: *START_NUTRS* = 2; *nSteps* = 10; haploid: $p_{axial}$ = 0.6, diploid: $p_{axial}$ = 0; *MF_STRENGTH* = 0.

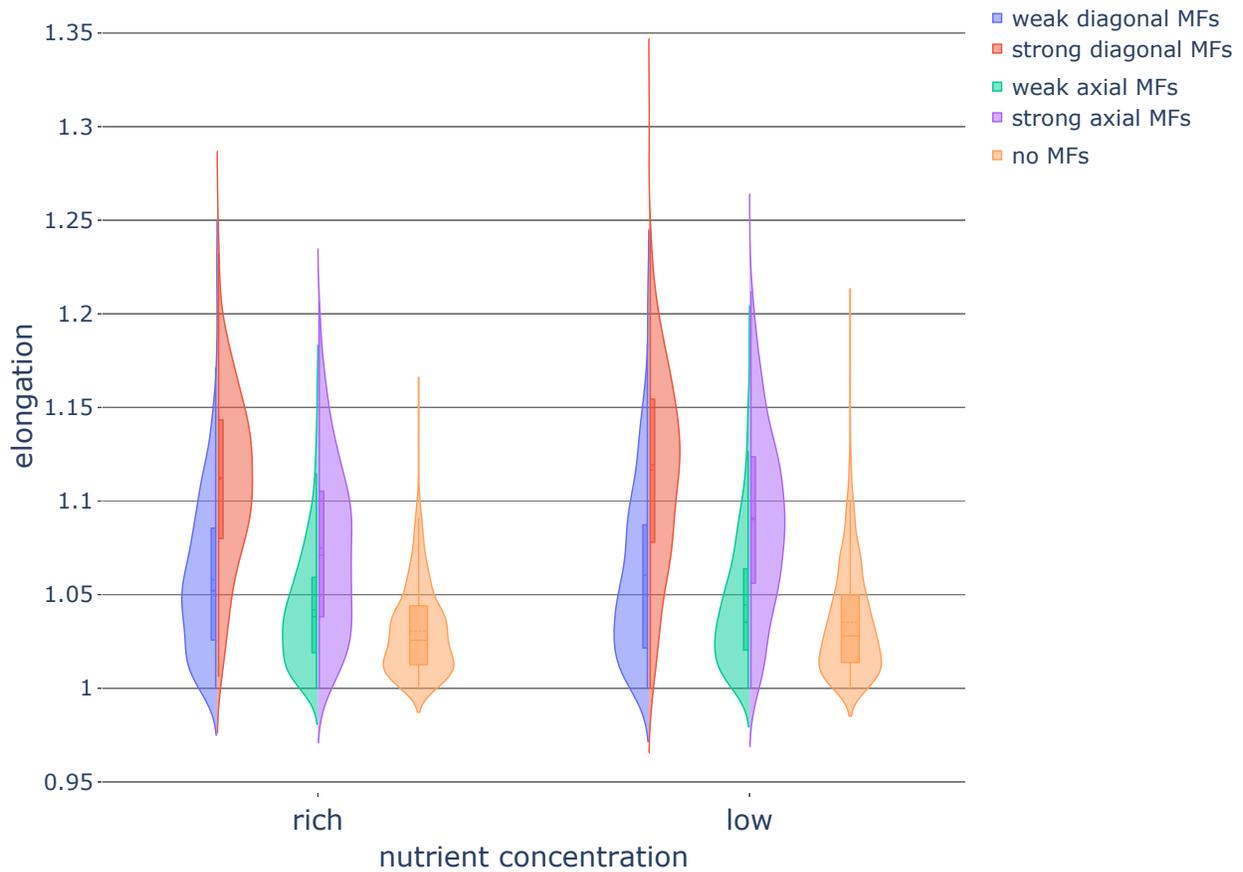

**Figure 16. Elongation of diploid colonies without pseudohyphal growth under different nutrient conditions and applied magnetic fields.** Violin plot of colony elongation under various nutrient concentrations and exposure to magnetic fields (MFs) of different strengths and directions. Box plots within the violin plots denote the median, quartiles, and outliers. Parameters were set as follows: rich-nutrient condition: *START_NUTRS* = 20, low-nutrient condition: *START_NUTRS* = 2; *nSteps* = 10; $p_{axial}$ = 0; no MFs: *MF_STRENGTH* = 0, weak MFs: *MF_STRENGTH* = 0.5, strong MFs: *MF_STRENGTH* = 1.

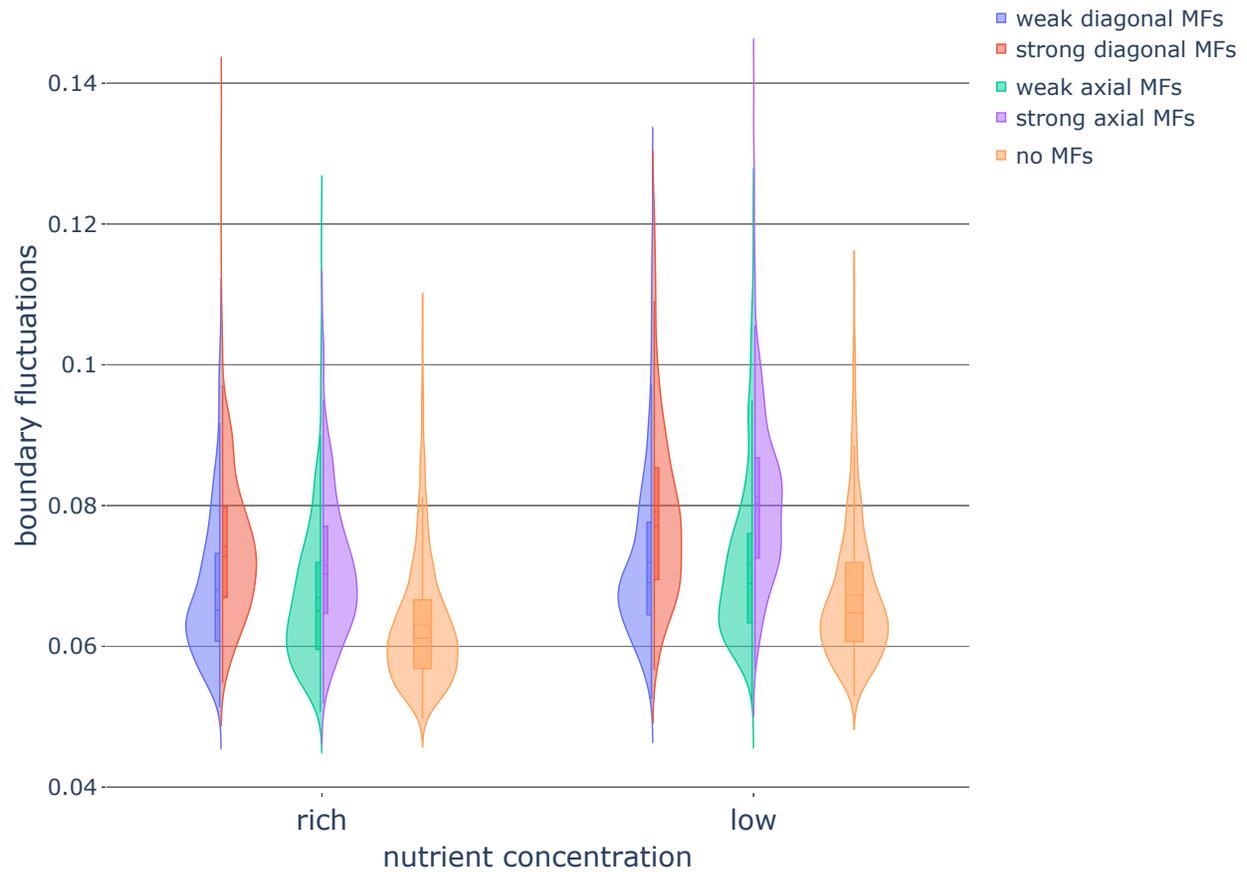

**Figure 17. Boundary fluctuations of diploid colonies without pseudohyphal growth under different nutrient and magnetic field conditions.** Violin plots of elongation of colonies in various nutrient conditions and exposed to magnetic fields (MFs) of different strengths and directions. Box plots within the violin plots denote the median, quartiles, and outliers. Parameters were set as follows: rich-nutrient condition: *START_NUTRS* = 20, low-nutrient condition: *START_NUTRS* = 2; *nSteps* = 10; $p_{axial}$ = 0; no MFs: *MF_STRENGTH* = 0, weak MFs: *MF_STRENGTH* = 0.5, strong MFs: *MF_STRENGTH* = 1.

**SUPPORTING INFORMATION (SI) APPENDIX:**

**Lattice-based simulation of the effects of nutrient concentration and magnetic field exposure on yeast colony growth and morphology**

Rebekah Hall[1] and Daniel A. Charlebois[2,3,*]

[1]Department of Mathematical and Statistical Sciences, University of Alberta, Edmonton, AB, T6G-2G1, Canada

[2]Department of Physics, University of Alberta, Edmonton, AB, T6G-2E1, Canada

[3]Department of Biological Sciences, University of Alberta, Edmonton, AB, T6G-2E9, Canada

*Correspondence:

Daniel Charlebois, Ph.D.
Department of Physics
Room 4-181, CCIS
11455 Saskatchewan Dr NW,
University of Alberta
Edmonton, AB, T6G-2E1
Canada
Tel: 780-492-3985
Email: dcharleb@ualberta.ca

**SUPPLEMENTAL FIGURES**

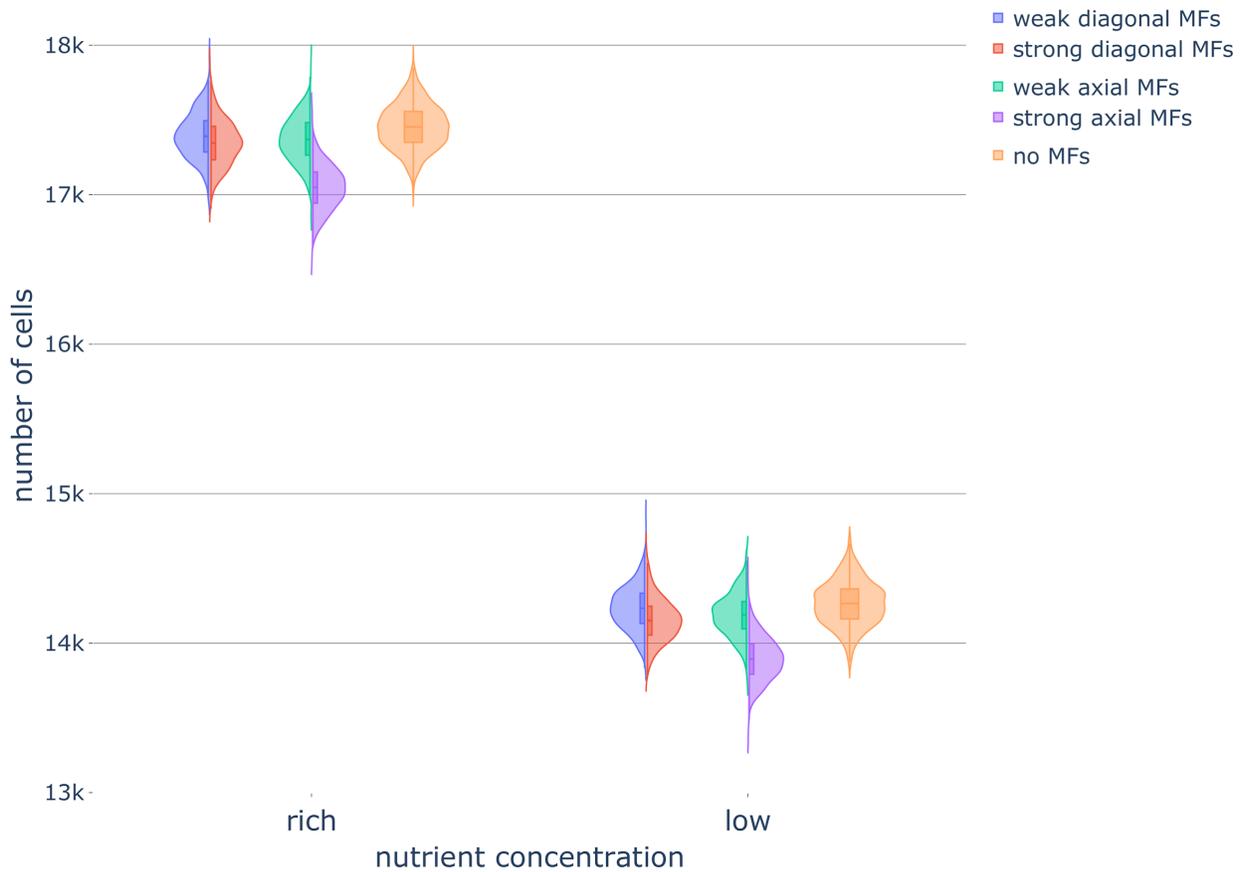

**Figure S1. Number of cells in haploid colonies under different nutrient and magnetic field conditions.** Violin plots of the number of cells in the yeast colony after 320,000 timesteps, when influenced by various magnetic field (MF) directions and strengths under rich-nutrient and low-nutrient conditions. Box plots within the violin plots denote the mean, quartiles, and outliers. Parameters were set as follows: rich-nutrient condition: *START_NUTRS* = 20, low-nutrient condition: *START_NUTRS* = 2; *nSteps* = 10; $p_{axial}$ = 0.6; no MFs: *MF_STRENGTH* = 0, weak MFs: *MF_STRENGTH* = 0.5, strong MFs: *MF_STRENGTH* = 1.

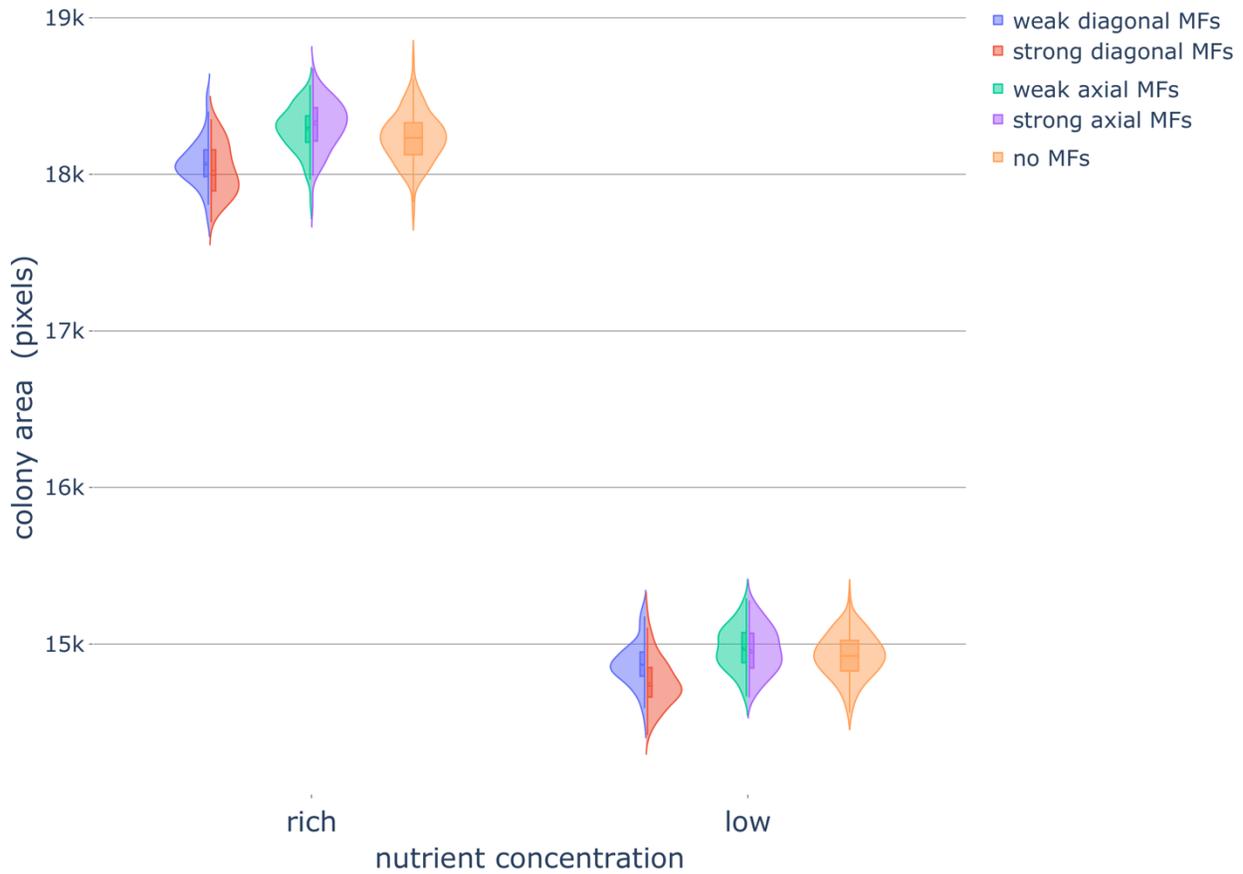

**Figure S2. Diploid colony area under different nutrient and magnetic field conditions.** Violin plots of yeast colony area after 320,000 timesteps, when influenced by various magnetic field (MF) directions and strengths under rich-nutrient and low-nutrient conditions. Box plots within the violin plots denote the mean, quartiles, and outliers. Parameters were set as follows: rich-nutrient condition: *START_NUTRS* = 20, low-nutrient condition: *START_NUTRS* = 2; *nSteps* = 10; $p_{axial}$ = 0; no MFs: *MF_STRENGTH* = 0, weak MFs*: MF_STRENGTH* = 0.5*,* strong MFs: *MF_STRENGTH* = 1.

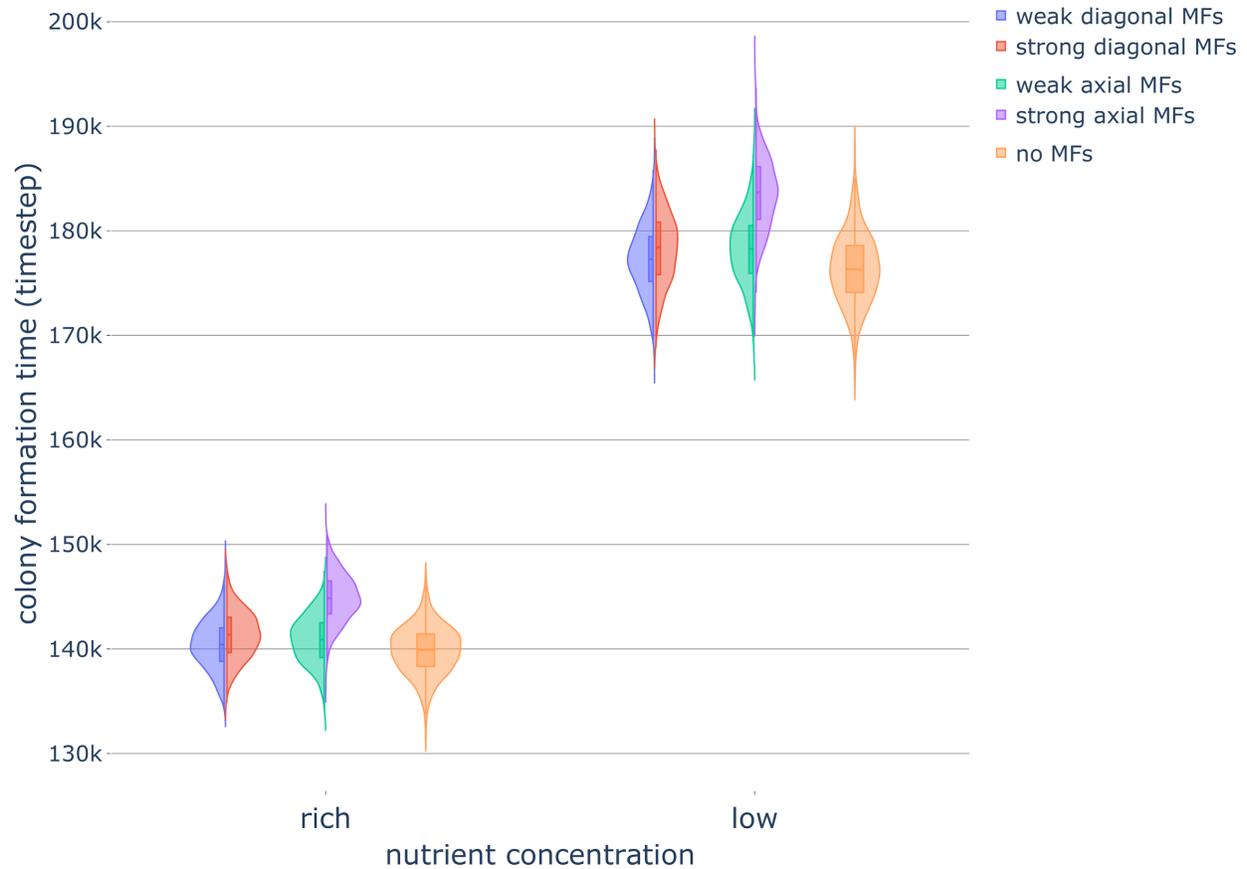

**Figure S3. Diploid colony formation time under different nutrient and magnetic field conditions.** Violin plots of the time to form a colony of 10,000 cells (starting from a single cell), when influenced by various magnetic field (MF) directions and strengths under low-nutrient and high-nutrient conditions. Box plots within the violin plots denote the median, quartiles, and outliers. Parameters were set as follows: rich-nutrient condition: *START_NUTRS* = 20, low-nutrient condition: *START_NUTRS* = 2; *nSteps* = 10; $p_{axial}$ = 0; no MFs: *MF_STRENGTH* = 0, weak MFs: *MF_STRENGTH* = 0.5, strong MFs: *MF_STRENGTH* = 1.

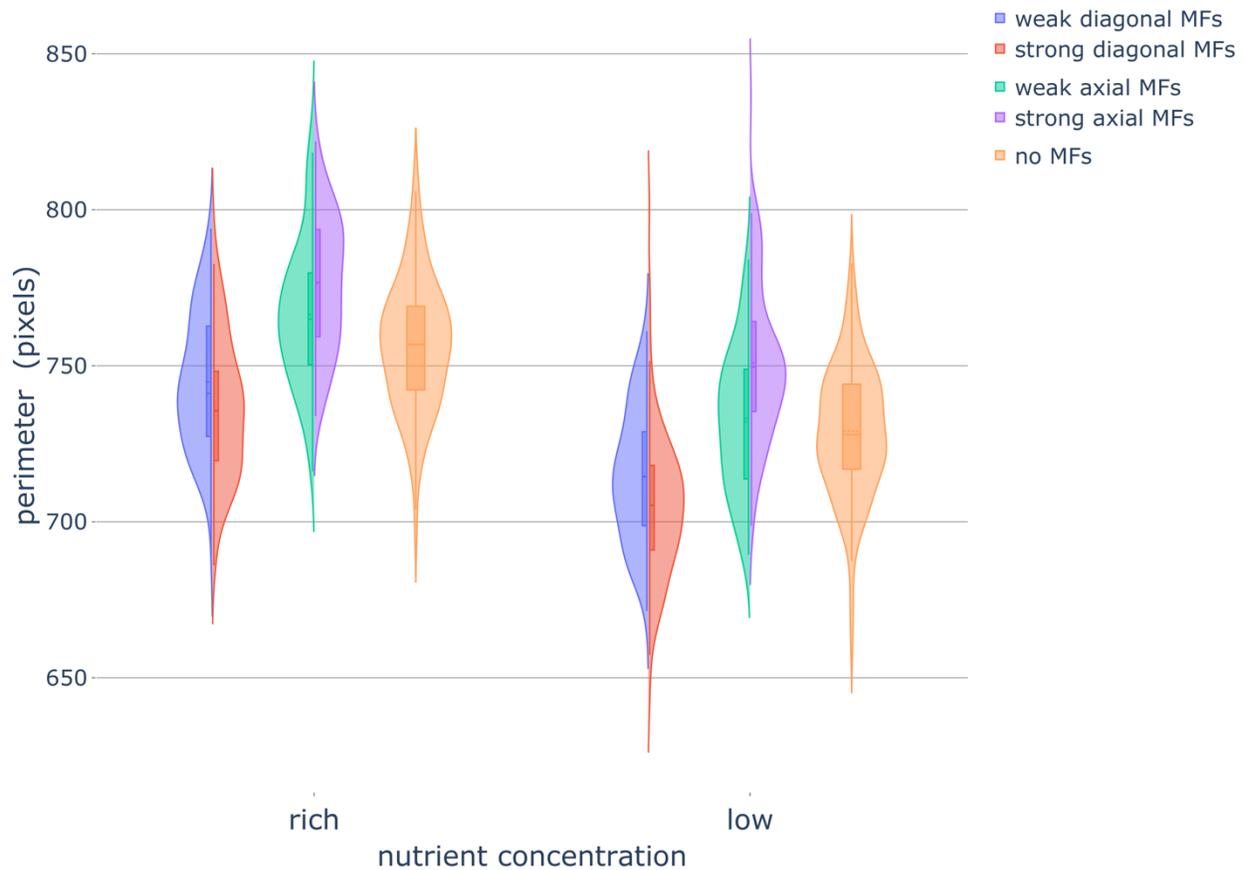

**Figure S4. Diploid colony perimeter under different nutrient and magnetic field conditions.** Violin plot of colony perimeters of colonies in various nutrient conditions, when exposed to magnetic fields (MFs) of different strengths and directions. Box plots within the violin plots denote the median, quartiles, and outliers. Parameters were set as follows: rich-nutrient condition: *START_NUTRS* = 20, low-nutrient condition: *START_NUTRS* = 2; *nSteps* = 10; $p_{axial} = 0$; no MFs: *MF_STRENGTH* = 0, weak MFs: *MF_STRENGTH* = 0.5, strong MFs: *MF_STRENGTH* = 1.

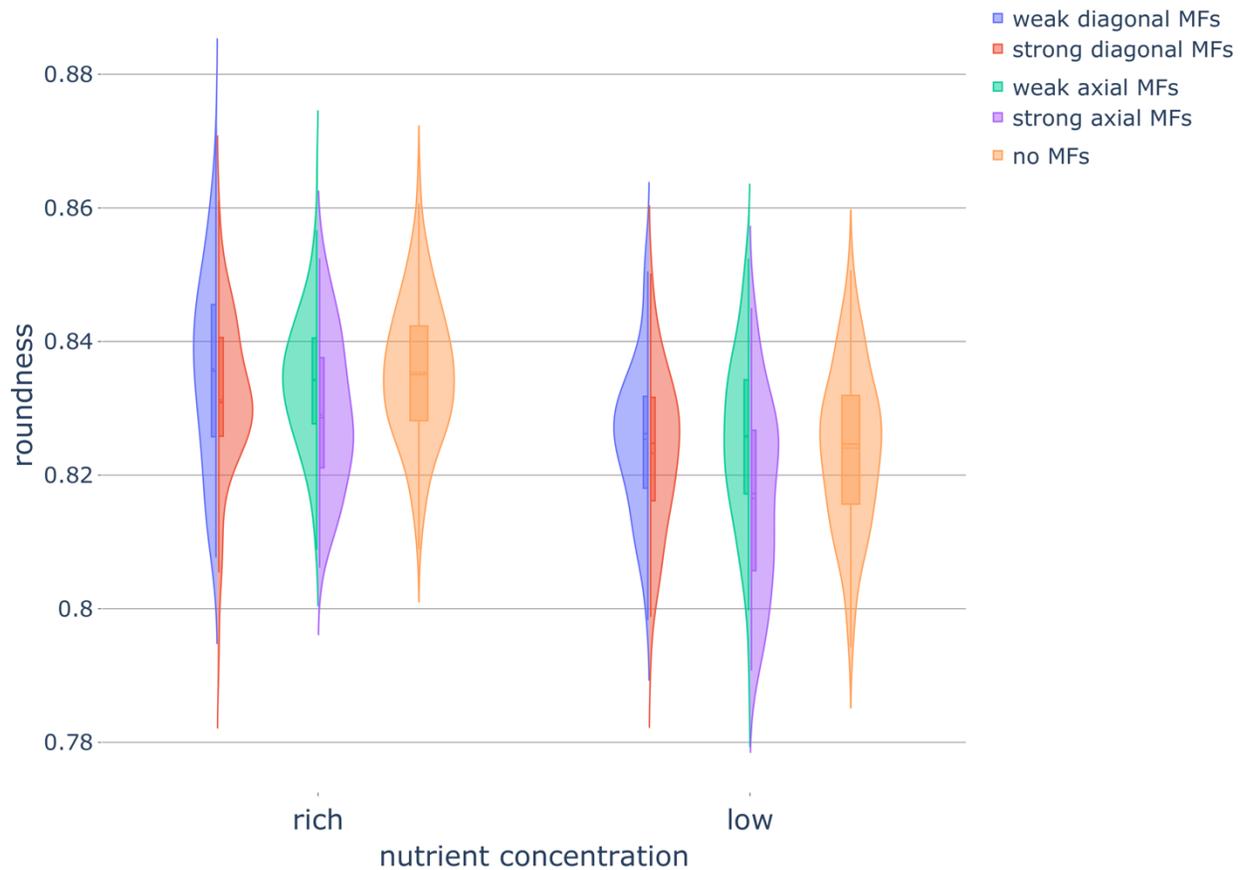

**Figure S5. Diploid colony roundness under different nutrient and magnetic field conditions.** Violin plots of colony roundness under different nutrient conditions and different magnetic field (MF) directions and strengths. Box plots within the violin plots denote the median, quartiles, and outliers. Parameters were set as follows: rich-nutrient condition: *START_NUTRS* = 20, low-nutrient condition: *START_NUTRS* = 2; *nSteps* = 10; $p_{axial} = 0$; no MFs: *MF_STRENGTH* = 0, weak MFs: *MF_STRENGTH* = 0.5, strong MFs: *MF_STRENGTH* = 1.

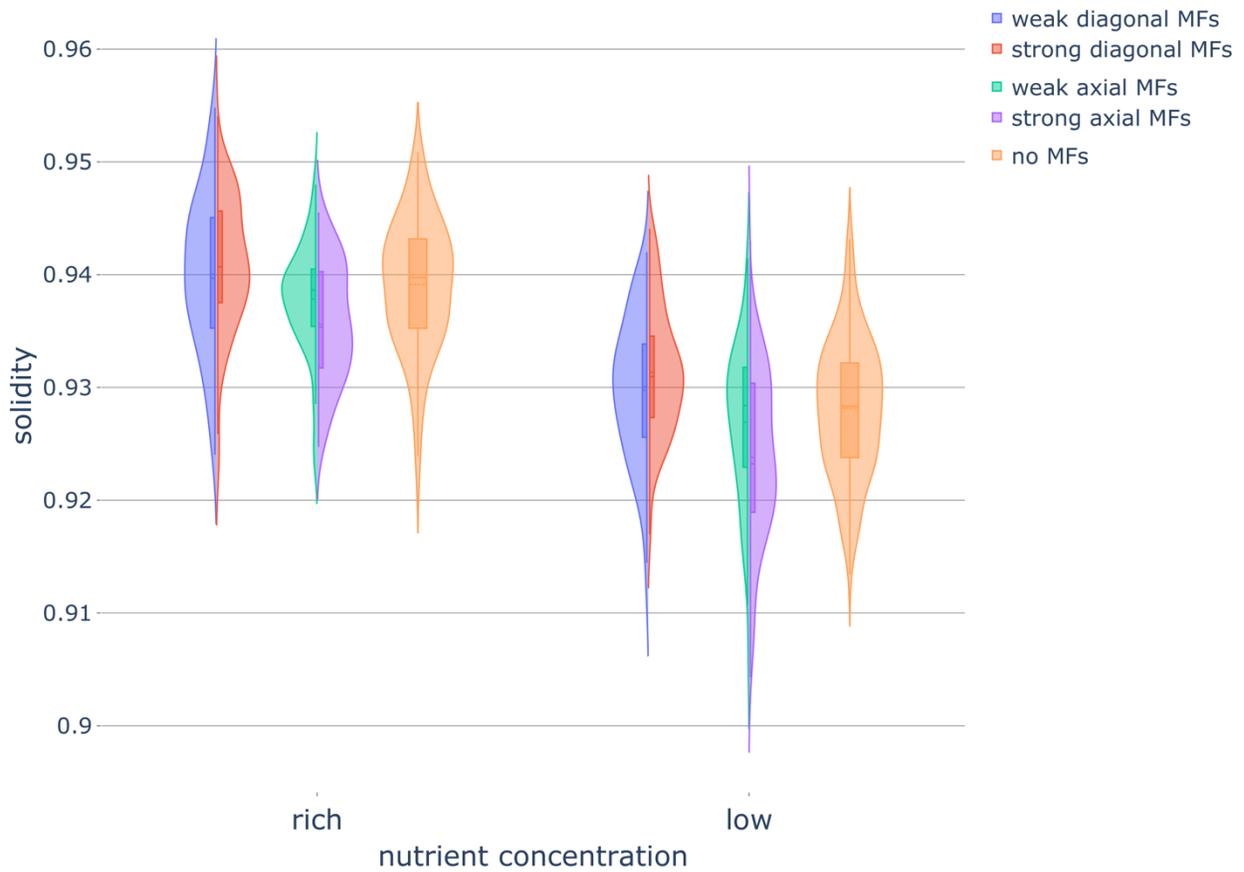

**Figure S6. Diploid colony solidity under different nutrient and magnetic field conditions.** Violin plot of colony solidity in various nutrient conditions while exposed to magnetic fields (MFs) of different strengths and directions. Box plots within the violin plots denote the median, quartiles, and outliers. Parameters were set as follows: rich-nutrient condition: *START_NUTRS* = 20, low-nutrient condition: *START_NUTRS* = 2; *nSteps* = 10; $p_{axial} = 0$; no MFs: *MF_STRENGTH* = 0, weak MFs: *MF_STRENGTH* = 0.5, strong MFs: *MF_STRENGTH* = 1.

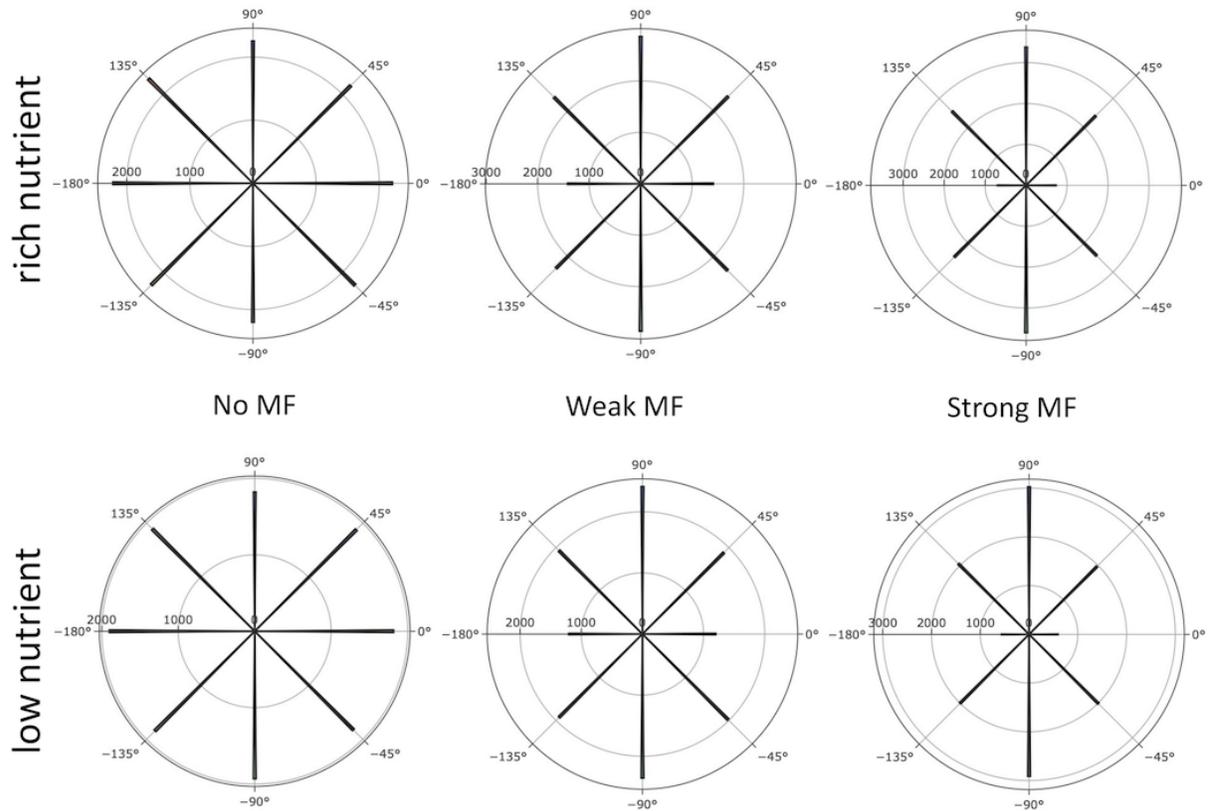

**Figure S7. Budding angle distributions for diploid colonies under different nutrient conditions of magnetic field strengths.** Plot of the frequency of the angle between the magnetic field direction and the mother-bud axis. Parameters were set as follows: rich-nutrient condition: *START_NUTRS* = 20, low-nutrient condition: *START_NUTRS* = 2; *nSteps* = 10; *MAGNETIC_FIELD* = [1 0]; no MFs: *MF_STRENGTH* = 0, weak MFs: *MF_STRENGTH* = 0.5, strong MFs: *MF_STRENGTH* = 1; *UNIPOLAR_ON* = false.